\documentclass[preprint2]{aastex}

\newcommand{\kmsmpc}{\kms\;{\rm Mpc}^{-1}}
\newcommand{\lya}{Ly$\alpha$\ }
\newcommand{\hkpc}{h^{-1}{\rm kpc}}
\newcommand{\hmpc}{h^{-1}{\rm Mpc}}

\newcommand{\sgal}{\sigma_{\rm gal}}
\newcommand{\sdm}{\sigma_{\rm DM}}
\newcommand{\kms}{\;{\rm km}\,{\rm s}^{-1}}
\newcommand{\bfit}{\beta_{\rm fit}}
\newcommand{\bspec}{\beta_{\rm spec}}
\newcommand{\lxunit}{{\rm erg\; s}^{-1}}
\newcommand{\entunit}{{\rm keV\; cm^{-2}}}
\newcommand{\fhot}{{f_{\rm hot}}}

\slugcomment{submitted to ApJ}

\shortauthors{Dav\'e, Katz, \& Weinberg}
\shorttitle{Simulated Group Scaling Relations}
\singlespace

\received{\today}
\begin{document}

\title{X-Ray Scaling Relations of Galaxy Groups in a
       Hydrodynamic Cosmological Simulation}

\author{Romeel Dav\'e \altaffilmark{1} }
\affil{Steward Observatory, University of Arizona, Tucson, AZ 85721}
\author{Neal Katz}
\affil{Astronomy Department, University of Massachusetts, Amherst, MA 01003}
\and
\author{David H. Weinberg}
\affil{Astronomy Department, Ohio State University, Columbus, OH 43210}

\altaffiltext{1}{Hubble Fellow, rad@as.arizona.edu}

\begin{abstract}

We examine the scalings of X-ray luminosity, temperature, and dark
matter or galaxy velocity dispersion for galaxy groups in a $\Lambda$CDM
cosmological simulation, which incorporates gravity, gas dynamics, radiative
cooling, and star formation, but no substantial non-gravitational heating.
In agreement with observations, the simulated $L_X-\sigma$ and $L_X-T_X$
relations are steeper than those predicted by adiabatic simulations or
self-similar models, with $L_X\propto \sigma^{4.4}$ and $L_X\propto
T_X^{2.6}$ for massive groups and significantly steeper relations
below a break at $\sigma \approx 180$~km/s ($T_X \approx 0.7$~keV).
The $T_X-\sigma$ relation is fairly close to the self-similar scaling
relation, with $T_X \propto \sigma^{1.75}$, provided that the velocity
dispersion is estimated from the dark matter or from $\ga 10$ galaxies.
The entropy of hot gas in low mass groups is higher than predicted by 
self-similar scaling or adiabatic simulations, and it agrees with observational
data that suggest an ``entropy floor.''
The steeper scalings of the luminosity relations
are driven by radiative cooling, which reduces the hot (X-ray emitting)
gas fraction from 50\% of the total baryons at $\sigma \approx 500$~km/s
to 20\% at $\sigma \approx 100$~km/s.  A secondary effect is that hot
gas in smaller systems is less clumpy, further driving down $L_X$.
A smaller volume simulation with eight times higher mass resolution predicts
nearly identical X-ray luminosities at a given group mass, demonstrating the
insensitivity of the predicted scaling relations to numerical resolution.
The higher resolution simulation predicts {\it higher} hot gas fractions at
a given group mass, and these predicted fractions are in excellent
agreement with available observations.  There remain some quantitative
discrepancies: the predicted mass scale of the $L_X-T_X$ and $L_X-\sigma$ 
breaks
is somewhat too low, and the luminosity-weighted temperatures are too high at a
given $\sigma$, probably because our simulated temperature profiles are flat
or rising towards small radii while observed profiles decline at 
$r \la 0.2R_{\rm vir}$.  We conclude that radiative cooling has an
important quantitative impact on group X-ray properties and can account
for many of the observed trends that have been interpreted as evidence
for non-gravitational heating.  Improved simulations and observations
are needed to understand the remaining discrepancies and to decide
the relative importance of cooling and non-gravitational heating in
determining X-ray scalings.
\end{abstract}

\section{Introduction}

Galaxy groups, i.e. bound systems having typically a few $L_\star$
galaxies, contain the majority of galaxies in the Universe, and hence seem
crucial to understanding the processes of galaxy formation.  However,
observational and theoretical studies have tended to focus either on
the smaller scale of individual galaxies or on the larger scale of rich
clusters.  Groups present an observational challenge, since projection
effects make them difficult to identify unambiguously in optical imaging
surveys and their X-ray emission is faint, emerging predominantly at
lower energies that are contaminated by foreground Galactic emission.
Sensitive X-ray instruments and group catalogs created from large
redshift surveys are transforming this observational situation, and
they have revealed puzzling departures in the group mass regime from
the scaling relations obeyed by rich clusters.  In this paper we use
a large hydrodynamic cosmological simulation to examine the predicted
X-ray scaling relations of galaxy groups, reaching to the mass regime
of poor clusters.  We adopt an inflationary cold dark matter scenario
with a cosmological constant ($\Lambda$CDM).

The canonical approach to understanding groups treats them as scaled
down versions of rich clusters.  The simplest model of a cluster, in
turn, is a sphere of Virial-temperature gas punctuated by old galaxies,
with perhaps a cooling flow onto the central cD galaxy.  In this model, the
gas cooling time is longer than a Hubble time everywhere except near
the center, and thus it is predicted that intracluster and intragroup
media should follow ``self-similar" scaling relations:  $T_X\propto
\sgal^2$, $L_X\propto T_X^2$, and $L_X\propto \sgal^4$, where $L_X$ is
the total X-ray luminosity, $T_X$ is the gas temperature (presumed to be
the halo Virial temperature), and $\sgal$ is the velocity dispersion of
cluster galaxies (presumed to trace the system's total mass).  The first
relation arises from equating the gas thermal energy to the galaxies'
kinetic energy (i.e. hydrostatic equilibrium), the second from combining
$L_X\propto M T_X^{1/2}$ for free-free emission (where $M$ is the system
mass), $M\propto \sgal^3$ from the Virial theorem, and the $T_X-\sgal$
relation, and the third from combining the previous two scaling relations
\citep[see e.g.][]{nav95}.

While observed properties of the most massive clusters follow these
relations \citep{all98,xu01}, the self-similar model breaks down as one
progresses to smaller systems, with observations indicating lower than
expected luminosities for a given temperature or velocity dispersion.
For 207 clusters observed with {\it Einstein}, \citet{whi97} found
$L_X\propto T_X^3$.  Extending such observations to groups has yielded
conflicting results; Mulchaey \& Zabludoff (1998, hereafter MZ98) found
$L_X\propto T_X^{2.8}$, consistent with clusters, while Helsdon \& Ponman
(2000, hereafter HP00) and \citet{xue00} found much steeper relations,
$L_X\propto T_X^{4.9-5.6}$.

In an influential paper, \citet[][hereafter PCN99]{pon99} calculated
the specific entropy of gas at 10\% of the Virial radius, and found
an ``entropy floor" of $\sim 100 h^{-1/3}$~keV~cm$^{-2}$ among their
smallest systems.  This entropy floor reduces the X-ray luminosity by
lowering the gas density progressively more in smaller systems, and hence
reproduces the steep $L_X-T_X$ relations seen in groups.  They suggested
that the entropy floor arises from energy injection by supernova winds,
which might concurrently explain the enrichment level of the intracluster
medium (ICM).  Though there are large uncertainties for the handful
of groups that demarcate the entropy floor, this result nevertheless
spawned numerous papers investigating the possible physical sources
of ``pre-heating,'' constraints on parameters of the pre-heating model
such as energy per baryon and energy injection redshift, and
implications of pre-heating for other group and cluster observables.
For instance, \citet{llo00} concluded that
a temperature increase of 0.3~keV per baryon is necessary to explain the
observed entropy floor, while \citet{toz01} considered a similar uniform energy
injection model and determined that the heat injection could be $\approx
0.1-1$~keV, depending on when and how it was injected.  

Uniform
heat injection across all baryons is ruled out by temperature measurements
of the \lya forest; these indicate $T\approx 2\times 10^{-3}$~keV for
most of the baryons at $z\sim 3$ \citep{sch99,bry00a}, and a significant
fraction of baryons at $z\sim 0$ are even cooler \citep{ric00,dav01}.
Thus, heat injection must be confined in and around halos.  Semi-analytic
models of this scenario require much greater energy injection, typically
$1-3$~keV per baryon within halos \citep{wu00,bow01} or entropy
injection of $\sim 300-400$~keV~cm$^{-2}$ \citep{bab01}, to reproduce
the PCN99 entropy floor, a result supported by adiabatic simulations of
clusters with pre-heating \citep{bia01,bor01}.  Supernovae seem unable
to provide this much energy, so another energy source such as AGN-driven
galactic winds is necessary \citep{val99,kra00,wu00,bow01,pip02}.  While there
is preliminary evidence for strong winds from Lyman break galaxies
\citep{pet01}, believed to evolve into cluster galaxies \citep{gov01},
the energy they inject seems unlikely to be as high as $\sim 1$~keV
per baryon.  The pre-heating model has also been invoked to explain
the faintness of the soft X-ray background \citep{pen99,wu01}, though
numerical simulations indicate that these constraints can probably
be satisfied without non-gravitational heating \citep{dav01b,phi01,cro01}.
In short, the pre-heating model is able to reproduce group and cluster
scaling relations, but only at the expense of invoking a mysterious
energy source that does not naturally arise from any firmly established 
physical
process of galaxy formation \citep[though see][for some interesting
hypotheses]{bab01}.  This is the essence of the ``ICM energy crisis"
described by \citet{toz01b}.

A fundamental assumption underlying the pre-heating model is that groups,
in the absence of such heating, would be self-similarly scaled-down
versions of large clusters.  If some process inherent in galaxy formation
breaks self-similarity, it may be that pre-heating is not required.
Growing evidence suggests that radiative cooling could be this process.
\citet{bry00} used an analytic model together with the observed dependence
of stellar mass fraction on cluster temperature to argue that the observed
scaling relations can be explained without non-gravitational heating.
\citet{voi01} further argued that cooling removes low-entropy gas during
the formation of cluster galaxies, leaving an effective entropy floor
of $\sim 100$~keV~cm$^{-2}$ for groups \citep{voi02}.
Hydrodynamic
simulations support these analytic arguments. \citet{mua01} found that
simulations without cooling followed the self-similar relations but that
the addition of cooling steepened and lowered the $L_X-T_X$ relation,
bringing it into much better agreement with observations.  \citet{tho01}
showed that adiabatic simulations underpredict the amplitude of the
$T_X-\sigma$ relation observed recently with {\it Chandra} but that
the addition of {\it either} radiative cooling {\it or} pre-heating
could resolve the discrepancy.  \citet{lew00} simulated a Virgo-sized
cluster with and without cooling and found that the inclusion of cooling and
subsequent star formation had an important impact on the X-ray emitting
gas throughout the entire cluster, not just in the central region
where the present-day cooling time is short.
These results clearly
point towards cooling as a non-negligible process in group and cluster
formation, and they motivate a more thorough investigation of cooling as an
alternative to pre-heating for explaining the observed scaling relations.

In this paper we analyze a cosmological hydrodynamic simulation
that incorporates cooling, star formation, and (weak) feedback.
Our simulation has enough dynamic range to probe the mass range from
poor groups to small clusters ($\sigma\approx 100-550$~km/s), while
resolving sub-$L_\star$ galaxies within a random cosmological volume.
We describe the simulation and analysis procedures in \S\ref{sec: sim}.
We present our simulated scaling relations in \S\ref{sec: scale} and
discuss the physical origin of the departures from self-similarity.
We examine the average profiles of various physical quantities in our
groups in \S\ref{sec: profiles}.  We go on to consider the possible
effects of metallicity, velocity bias, and surface brightness thresholds
in \S\ref{sec: obseff}.  These effects introduce some uncertainty in
our predictions of observable scaling relations; we discuss our ``best
guess'' predictions and compare them to observations in \S\ref{sec: obs}.
Section~\ref{sec: concl} presents our conclusions.  In summary, we find
that radiative cooling --- in particular, the greater efficiency of
cooling in smaller systems --- can produce departures from self-similar
scaling that quantitatively mimic those of a pre-heating model.  Thus,
observed scaling relations do not necessarily imply that groups have
experienced substantial non-gravitational heating or entropy injection.

\section{Simulation and Group Identification}\label{sec: sim}

We simulate a $50\hmpc$ random volume with a $7\hkpc$ (equivalent Plummer,
comoving) gravitational softening length, assuming a $\Lambda$CDM
cosmological model with $\Omega_m=0.4$, $\Omega_\Lambda=0.6$, $h \equiv
H_0/100\;\kmsmpc = 0.65$, $n=0.95$, and $\sigma_8=0.8$.  We employ
$144^3$ dark matter and $144^3$ gas particles, yielding particle masses 
of $m_{\rm bary}=8.5\times 10^8 M_\odot$ and $m_{\rm dark}=7.2\times
10^9 M_\odot$.  The input physics includes cooling, star formation,
and thermal feedback.  Because the feedback energy is deposited locally
in the dense gas giving rise to star formation, it is usually radiated
away quickly and produces little long-term heating of the surrounding
medium.  Other than this feedback, we do not inject any non-gravitational
heat or entropy.  The simulation was evolved from $z=49\rightarrow 0$
using Parallel TreeSPH \citep{dav97}.  A detailed discussion of the
algorithms, including the star formation and feedback prescriptions,
is given by \citet{kat96}.  Relative to the simulation of \cite{mua01},
who investigate similar issues with a similar numerical approach, our
simulation has a mass resolution that is higher by about a factor of four
and gravitational force resolution that is higher by a factor of 4--7
(depending on redshift), but a simulation volume smaller by a factor
of eight.  Our simulation is therefore better suited to the study of
groups, but not as good for the study of rarer, more massive clusters.

We identify galaxies using SKID\footnote{Spline
Kernel Interpolative DENMAX, publicly available at {\tt
http://www-hpcc.astro.washington.edu/tools/skid.html}.}.  Our
60-particle galaxy mass resolution limit ($5\times 10^{10} M_\odot$)
corresponds to $\approx L_\star/4$, based on the equivalent number density
of observed galaxies in the Sloan Digital Sky Survey (\citealt{bla01};
for further discussion of the resolution limit and completeness
threshold of the simulated galaxy sample see \citealt{mur02}).
We identify bound systems using a spherical-overdensity (SO) criterion on
friends-of-friends halos, as described by \citet{gar01}.  In this method,
the most bound particle within a friends-of-friends halo is found, and
the system extends radially outwards until the enclosed average density
reaches the Virial density threshold (282 times the mean density for
our cosmology).  SO systems that contain three or more galaxies above
our resolution limit are identified as ``groups."  We find 128 groups
at $z=0$, spanning a mass range of $10^{12.1}-10^{14.5}M_\odot$, with
up to 42 member galaxies.  Note that all of these groups are bound;
we identify groups in three dimensions, so our catalog does not contain
chance line-of-sight projections.

To study the effects of numerical resolution, we also analyze a
smaller volume, higher resolution simulation having the same cosmological
model; we will refer to this as our ``high-resolution" simulation.  This 
simulation
has $2\times 128^3$ particles in a $22.222\hmpc$ volume, with $3.5\hkpc$
(equivalent Plummer, comoving) softening, yielding particle masses of
$m_{\rm bary}=1.05\times 10^8 M_\odot$ and $m_{\rm dark}=8.8\times 10^8
M_\odot$.  We find 52 groups in this simulation at $z=0$.  By chance,
this volume contains one anomalously large ($10^{14.3}M_\odot$) group
containing 217 galaxies above our 60-particle galaxy mass limit of
$6.3\times 10^9 M_\odot$, but the other 51 groups probe the expected
mass range down to $10^{11.2}M_\odot$.

The velocity dispersion of a group can be calculated from its dark
matter ($\sdm$) or its member galaxies ($\sgal$).  We quote 1-D
velocity dispersions, obtained by dividing the 3-D velocity dispersion
by $\sqrt{3}$.  We find that $\sdm$ correlates tightly with group mass
while $\sgal$ shows systematic deviations from $\sdm$ for smaller groups,
as discussed below in \S\ref{sec: sigma}.

To calculate X-ray emission, we use the latest plasma code of
\citet{ray77}.  The interface to simulations via the TIPSY\footnote{
Publicly available at {\tt
http://www-hpcc.astro.washington.edu/tools/tipsy/tipsy.html}.} package
was kindly provided by G. Lewis and C. Murali.  We calculate particle
luminosities in the $0.5-2$~keV bandpass that is most commonly employed to
study group X-ray properties using {\it ROSAT}.  Unless otherwise noted,
the X-ray temperature is calculated from the average luminosity-weighted
temperature of group particles; we will discuss some issues relating to
this in \S\ref{sec: bspec}.

Applying the usual SPH formalism to obtain gas densities for calculating
luminosities results in a gross overestimation of the X-ray emission.
The usual SPH algorithm does a poor job of resolving the two-phase
interface between the intragroup medium and gas in galaxies.  As a
consequence, the density of a hot intragroup gas particle
is greatly overestimated
when it happens to lie very close to a cold, dense clump, producing large
and unphysical X-ray emissivities for some of the particles.  
To correct for this effect, we follow \citet{pea00} and explicitly
decouple the hot and cold phases by recalculating gas densities using
only hot particles with $T>10^{5}$K.  Unlike Pearce et al., however,
we only implement decoupling at the post-processing stage, not during
the simulation itself.  Our results are insensitive to the choice of
temperature threshold, so long as it is below $T\la 10^{6}$K, since
virtually all gas particles bound in groups that have not cooled into
galaxies have temperatures higher than this threshold.  \citet{cro01}
performed extensive tests of this approach, and found that it is indeed
insensitive to the exact temperature threshold, behaves as expected with
resolution, and reproduces the correct X-ray luminosity in analytic cases.
The number of particles affected is small, but the total
correction factors are large \citep[typically a few to ten; see][]{cro01}, so
our luminosity predictions must be read with this caution in mind.
Additionally,
since we do not explicitly decouple the hot and cold phases while evolving
the simulation, we probably overpredict the amount of cooled baryons,
by an uncertain amount.  
(The cold phase is ultimately ``decoupled'' from the hot phase by conversion
into collisionless stars, but this conversion is not instantaneous.)
\cite{spr01} suggest that an alternative
formulation of the SPH equations can resolve this ``decoupling problem''
in a self-consistent manner, and we will investigate this possibility
in future simulations.

\section{Testing Self-Similarity}\label{sec: scale}

\subsection{Simulated Group Scaling Relations}

We begin by examining whether our simulated groups follow the self-similar
scaling relations expected for systems with negligible non-gravitational
heating.  To do this, we compute X-ray luminosities using an
intragroup metallicity of zero, we sum the luminosity of all gas in
the group out to its Virial radius, and we take the group's velocity
dispersion to be that of the dark matter.  If group formation were well
described by non-radiative physics (a.k.a. ``adiabatic'' physics, 
though gravitational shock heating is also included),
then this approach should closely reproduce the
self-similar scaling relations (see, e.g., \citealt{owe98,mua01}).

\begin{figure}
\plotone{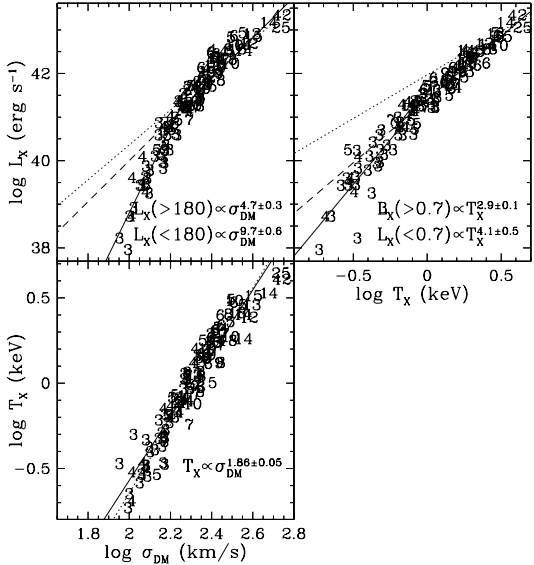}
\caption{Scaling relations of the simulated galaxy groups, assuming
zero metallicity.
The upper left panel shows $L_X-\sdm$, upper right $L_X-T_X$, and lower
left $T_X-\sdm$.  The symbols indicate the number of galaxies
in each group.  The best-fit relations are listed in the lower right.
The solid line shows the best fit, having a break at the value indicated
in the legend (except in the lower left panel, where a single power
law fit is acceptable).
The dashed line is a continuation of the slope above the
break to lower values of $\sdm$ or $T_X$.  The dotted lines show predictions
from self-similar scaling, normalized to our largest groups.
These scaling relations may be compared to, e.g., Figures~4-6 of
\citet{mul00}; we will present a more detailed comparison to observations
in \S\ref{sec: obs}.  In no regime do our simulated
luminosity scaling relations follow those predicted by self-similar models
and by adiabatic numerical simulations.
\label{fig: scaling}}
\end{figure}

Figure~\ref{fig: scaling}, upper left, shows the $L_X-\sdm$ relation
for our groups.  The plot symbols indicate the number of galaxies
in each group.  A break is clearly evident, occurring at $\sdm\approx
180$~km/s; the fit above the break is extended to lower $\sdm$ as the
dashed line, and it is clearly discrepant with these smaller systems.
The dotted line shows the self-similar scaling relation, $L_x \propto
\sdm^4$, normalized to our largest groups.  The break value is determined
by requiring continuity between the fits above and below the break.
Best fit power law relations above and below this break value are listed
in the lower right.  Even above the break, the $L_x-\sdm$ relation is
steeper than predicted by the self-similar model, with $L_X\propto
\sdm^{4.7\pm 0.3}$.  Below the break the relation is very steep,
$L_X\propto \sdm^{9.7\pm 0.6}$.  The fits are unweighted, to follow
the procedure most commonly used by observers.  We list our scaling
relation fits in Table~1, including the scaling relations observed by
MZ98 and HP00.  We defer a comparison with observations until \S\ref{sec:
obs}, where we include various observational effects in our analysis;
here we simply note that the observations show a similar qualitative
trend in which luminosities drop faster with decreasing group mass than
the self-similar model predicts.

Figure~\ref{fig: scaling}, upper right, shows the $L_X-T_X$
relation.  Here again a break is evident, though less obviously so,
at $T\approx0.7$~keV.  Above the break, the relation is $L_X\propto
T_X^{2.9\pm 0.1}$, clearly discrepant with self-similarity ($L_X
\propto T_x^2$, dotted line), while below the break it steepens to
$L_X\propto T_X^{4.1\pm 0.5}$.  A single power-law fit yields a slope of
$3.4\pm 0.1$.  Here the break value is somewhat more ambiguous, but even
with the break placed at $T=2$~keV (in which case the fits are clearly
discontinuous), the slope above the break is still 2.6, and inconsistent
with self-similarity.  Though a double power law is an adequate fit,
there is an overall trend for a continuous steepening of the relation
to lower temperatures.

Our luminosity scaling relations demonstrate two points.  The first is
that a break occurs at all, 
when the simulation incorporates no physics ``tuned'' to pick out
this mass or temperature scale.
The second is that even above the break, the slope shows a significant
departure from self-similarity.  In no regime do our groups follow the
expected self-similar relations, suggesting that self-similarity is
an inappropriate model for relating clusters and groups.  This strong
departure from self-similarity is our most important result; the rest
of the paper will be devoted to understanding its origin and comparing
our predictions to observed group properties.

The lower left panel of Figure~\ref{fig: scaling} shows the $T_X-\sdm$
relation.  In contrast to the luminosity relations, this relation shows
no break, and the scaling, $T\propto \sdm^{1.86\pm 0.05}$, is only
slightly off from the self-similar prediction ($T_X \propto \sdm^2$,
dotted line).  This agreement suggests that the X-ray temperature
and dark matter velocity dispersion are both determined mainly by the
gravitational potential, as expected for systems that are Virialized
and in hydrostatic equilibrium.  The slight excess heating that makes
the relation shallower than $T_X \propto\sdm^2$ may be coming from shock
heating of gas on filaments before accretion into the group potential,
which can heat gas to $\sim 10^6$~K \citep{dav01b}.  Supernova feedback
as implemented in this simulation is not expected to add any significant
heat to the intergalactic gas.  In any case, the amount of heating indicated
in this plot is far from sufficient to explain the departures from
self-similarity in the $L_X-T_X$ relation.  Moreover, since there is no
evidence of a break in $T_X-\sdm$, it is clear that some other physical
process is at work in the luminosity scaling relations.

\begin{deluxetable}{ccc|cc|cc|c}
\footnotesize
\tablecaption{Simulated Group Scaling Relations.}
\tablewidth{0pt}
\tablehead{
\colhead{Z\tablenotemark{a}} &
\colhead{SB\tablenotemark{b}} &
\colhead{$\sigma$} &
\colhead{$\bigl[\frac{d\log{L_X}}{d\log{\sigma}}\bigr]_{>180}$} &
\colhead{$\bigl[\frac{d\log{L_X}}{d\log{\sigma}}\bigr]_{<180}$} &
\colhead{$\bigl[\frac{d\log{L_X}}{d\log{T_X}}\bigr]_{>0.7}$} &
\colhead{$\bigl[\frac{d\log{L_X}}{d\log{T_X}}\bigr]_{<0.7}$} &
\colhead{$\bigl[\frac{d\log{T_X}}{d\log{\sigma}}\bigr]$}
}
\startdata
0 & none & $\sdm$ & $4.7\pm 0.3$ & $9.7\pm 0.6$ & $2.9\pm 0.1$ & $4.1\pm 0.5$ 
& $1.86\pm 0.05$ \\
eq.~\ref{eqn: metal} & none & $\sdm$ & $4.3\pm 0.2$ & $9.5\pm 0.6$ & $2.5\pm 
0.2$ & $4.6\pm 0.3$ & $1.69\pm 0.05$ \\
$0.3Z_\odot$ & none & $\sdm$ & $3.8\pm 0.2$ & $7.7\pm 0.5$ & $2.2\pm 0.2$ & 
$4.0\pm 0.3$ & $1.77\pm 0.05$ \\
0 & eq.~\ref{eqn: rx} & $\sdm$ & $5.0\pm 0.3$ & $11.9\pm 0.9$ & $3.4\pm 0.2$ & 
$4.8\pm 0.4$ & $1.88\pm 0.06$ \\
0 & none & $\sgal$ & $3.9\pm 0.4$ & $3.9\pm 0.9$ & $2.9\pm 0.1$ & $4.1\pm 0.5$ 
& $1.14\pm 0.07$ \\
eq.~\ref{eqn: metal} & eq.~\ref{eqn: rx} & $\sdm$ & $4.4\pm 0.2$ & $11.7\pm 
0.9$ & $2.6\pm 0.1$ & $4.8\pm 0.4$ & $1.75\pm 0.05$ \\
MZ98 & - & - & $4.3\pm 0.4$ & - & $2.8\pm 0.1$ & - & $2.2\pm 0.9$ \\
HP00 & - & - & $4.5\pm 1.1$ & - & $4.9\pm 0.8$ & - & $0.9\pm 0.2$
\enddata
\tablenotetext{a}{Assumed metallicity.}
\tablenotetext{b}{Surface brightness correction.}
\end{deluxetable}

\subsection{Physical Origin of the Simulated Scaling Relations}
\label{sec: baryfrac}

Why does our simulation yield lower-than-expected luminosities as one goes
to smaller systems?  The origin of this effect does not appear to be some
heat source that adds pressure support and lowers the central density
at a given Virial mass, nor does it appear that 
the low mass systems have lower luminosities because of substantially
sub-Virial temperatures.
Either of these scenarios would 
require a feature in the $T_x-\sigma$ relation close to the break 
at $\sigma \approx$ 180~km/s in the $L_X-\sigma$ relation, and none is seen.
Alternative explanations are that the X-ray emitting gas density is
lowered by removal of gas from this phase, or that the density structure
of the gas is changing systematically with group mass.  Both of these
effects are, in fact, occurring in our simulation, and together they
account for the departure from the self-similar scaling relations.

\begin{figure}
\plotone{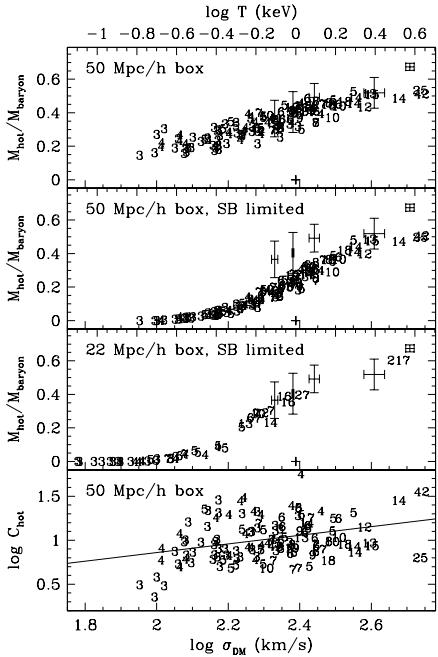}
\caption{{\it Top panel:} Ratio of hot ($T>10^{5}$K) to total baryons
as a function of group velocity dispersion.  The top axis legend 
indicates the approximate group temperature, obtained by using the $T_X-\sdm$
scaling relation found in Figure~1.  The data points show binned 
observational values
as described in the text.  {\it Second panel:} Same as top panel, but
results have been computed only out to an ``observable" fraction of
the Virial radius, as given by equation~(\ref{eqn: rx}).  {\it Third
panel:} Same as second panel, but based on the high-resolution,
smaller volume simulation.
{\it Fourth panel:} ``Clumping factor'' (eq.~\ref{eqn: clump}) 
of hot gas in each group, from the large-volume simulation.
\label{fig: baryfrac}}
\end{figure}

The top panel of Figure~\ref{fig: baryfrac} shows the fraction 
$\fhot \equiv M_{\rm hot}/M_{\rm baryon}$ of each group's total
baryons that are in the hot, X-ray emitting phase.  There is a clear trend
that smaller systems have less X-ray emitting gas.  Our largest systems
have slightly more than half of their baryons in a hot phase, while our
smallest systems have less than 20\%.  The trend steepens slightly to
smaller systems.  The steepening is visually exacerbated when one
plots $\fhot^2$ on a logarithmic plot, 
hence the appearance of a ``break" in the $L_X-T_X$
and $L_X-\sdm$ relations.

The data points in Figure~\ref{fig: baryfrac} are binned from
a compilation of observations of 30 X-ray groups and clusters
with temperatures ranging from 0.8-4.5~keV.  The lowest 17 are from
\citet{mul96}, the next five are from \cite{hwa99}, and the highest eight
are $T<4.5$~keV Abell clusters from \citet{cir97}.  The error bars shown
indicate the error in the mean computed from the scatter among the data
points in each bin.  In each case, we take the ratio of the gaseous to
gaseous+stellar mass estimates.  The observations do show a trend of
decreasing hot gas fraction with decreasing group mass.  Unfortunately,
current data do not extend to the mass range of the lowest mass groups
in our simulation, where the reduction in hot gas is largest.

In the top panel of Figure~\ref{fig: baryfrac}, there appears to be
remarkably good agreement between the simulated and observed hot gas
fractions.  Unfortunately, this agreement is somewhat misleading.  
As \citet{bal01}
have argued, X-ray observations of groups do not extend nearly as far out,
relative to the virial radius, as they do in clusters, and since the hot
gas is distributed more diffusely than the cold gas, the observed hot
gas fraction in smaller groups is biased low.  While it is difficult
to correct the observations, it is possible to model this bias in our
simulated groups to enable a fairer comparison.  To do so, we recalculate
the baryon fractions only within a radius given by
\begin{equation}\label{eqn: rx}
\frac{R_X}{R_{\rm vir}} = 1.09 \log{\left(\frac{\sdm}{\rm km/s}\right)} - 2.11,
\end{equation}
where $R_X$ is the extent of the X-ray emission. 
This relation is obtained by fitting the data in Figure~3 of \citet{mul00}
[for which we obtain $R_X/R_{\rm vir}=(\log{T}+0.7)/1.7$]
and using the $T_X-\sdm$ relation from Figure~\ref{fig: scaling}.
We take $R_{\rm vir}$ to be the maximum extent of the group.

The result of applying equation~(\ref{eqn: rx}) is shown in the second
panel of Figure~\ref{fig: baryfrac}.  The predicted hot fraction is indeed
smaller than in the top panel, increasingly so for smaller systems.  Thus,
in a fairer comparison with data, our simulation appears to ``overcool''
baryons in groups.  As we have discussed elsewhere \citep{dav01b,mur02},
the simulation also appears to ``overcool'' baryons globally --- the
fraction of baryons converted to stars is
24\%, which is high compared to estimates from the
observed luminosity function \citep{bla01,col01}, though it is consistent
with estimates from the extragalactic background light \citep{ber01}.

\citet{pea00} and \citet{bal01} have argued that a higher resolution
simulation should produce substantially more cooled gas.
We can test this expectation directly using our high-resolution simulation.
The third panel of Figure~\ref{fig: baryfrac} shows the hot fraction for the 52
groups identified in this simulation, including the surface brightness
cut described above.  Interestingly, the trend is opposite to that
argued previously: there is actually somewhat {\it more} hot gas in the higher
resolution simulation.  We believe that this increase in hot gas fraction with
increasing numerical resolution, i.e. a decrease in galaxy masses, is caused by
the difficulty in resolving the interface of two such distinct phases.
This causes cooling rates in the
immediate vicinity of galaxies to be overestimated (see \S\ref{sec: sim}),
and affects a smaller fraction of particles in the higher resolution 
simulation.
The hot gas fractions in our high-resolution simulation, including surface
brightness bias, are in good agreement with the observations, though the
range of overlap is still limited, and we do not have a still higher
resolution simulation with which to demonstrate convergence of the numerical 
prediction.  The {\it global} cold fraction in the 
high-resolution simulation is slightly higher than that of the large-volume 
simulation,
with 26\% of the baryons converted to stars, since the simulation
resolves cooling and star formation further down the mass function.
An estimate of the total stellar mass fraction, obtained by combining together
several simulations of varying volume and resolution is presented in Fardal
et al. (2002 in preparation).


The simple uniform removal of hot gas is not sufficient
to fully explain the reduced luminosities of lower mass groups.
Consider groups with $T_X
\approx 3$~keV ($\log T \approx 0.5$), having hot gas fractions
$\sim 0.5$, and those with $T_X \approx 0.3$~keV, having hot gas
fractions $\sim 0.2$.  Relative to a model with constant gas fraction,
the low mass groups in our simulation have hot gas densities lower by a
factor of 2.5.  Since $L_X \propto n_{\rm hot}^2$, this reduction should
depress the luminosities of the low mass groups by a factor of 6.25.
However, the gap between our best-fit relation and the self-similar
scaling (solid and dotted lines of Figure~\ref{fig: scaling}, at $\log
T_X = -0.5$) is approximately a factor of 25.  This shows that the
{\it uniform} removal of gas accounts for somewhat more than half
(logarithmically) of the reduction in luminosity.

The remaining factor arises because hot gas is less clumped in the smaller
systems.  The bottom panel of Figure~\ref{fig: baryfrac} shows the
clumping factor
\begin{equation}\label{eqn: clump}
C_{\rm hot} = \frac{\langle \rho_{\rm hot}^2\rangle}
                   {\langle \rho_{\rm hot}  \rangle^2} = 
\frac{\sum_{i=1}^{N_{\rm hot}} \rho_i}{N_{\rm hot} \bar{\rho}_{\rm hot}},
\end{equation}
for each of our systems.
The $\langle ..\rangle$ in the first equality represent volume averages,
and the second equality expresses this ratio as a sum over 
(equal mass) SPH particles, where $\rho_i$ is the SPH-estimated density
of particle $i$, $N_{\rm hot}$ is the number of hot
gas particles in the group, and $\bar{\rho}_{\rm hot}$ is the mean density of
hot gas, namely the total hot gas mass of the group divided by the volume
out to the Virial radius.  If the hot gas in these groups were isothermal,
then this clumping factor would be directly
proportional to the luminosity \citep[cf.][]{evr96},  
since the volume emissivity is proportional to $\rho_{\rm hot}^2$
throughout the group.
If $C_{\rm hot}$ were the same in high and low mass groups, then
their relative luminosities would scale in proportion to $f_{\rm hot}^2$
(plus temperature factors, but the group temperatures {\it do} follow
the self-similar scaling to a good approximation).
The term ``clumping factor'' should be applied with some caution,
since the globally averaged quantity defined in equation~(\ref{eqn: clump})
could be different in two groups that have smooth gas distributions
but different radial density profiles $\rho_{\rm hot}(r)$.

Figure~\ref{fig: baryfrac} shows that $C_{\rm hot}$ drops by a factor
of $\sim 3$ from 3~keV systems to 0.3~keV systems (according to the
best-fit relation shown as the solid line), though the scatter is large.
This indicates that clumping is responsible for most of the remaining
drop in luminosity, after accounting for changes in $\fhot$.
The physical reason for the drop in $C_{\rm hot}$
is not obvious.  One possibility is that larger systems are younger and
hence less dynamically relaxed, and therefore contain more substructure.
Another possibility is suggested by the analytic model of \citet{bry00},
who also predicts a luminosity drop of a similar factor due solely to
reduced ``clumping'' in smaller systems (compare his dot-dashed and solid
lines in his Figure~3).  In his case, this difference 
arises because cooling more
efficiently removes hot gas in the central regions of smaller systems.
As we will show in \S\ref{sec: profiles}, we do not see a drop in
the {\it mean} density in the inner regions of smaller systems, however
the differences in $C_{\rm hot}$ do arise mainly in the inner regions.
Regardless of cause, it is the combination of lowered subclumping and
an overall drop in the hot fraction that is responsible for the lowered
luminosities in smaller groups.

\begin{figure}
\plotone{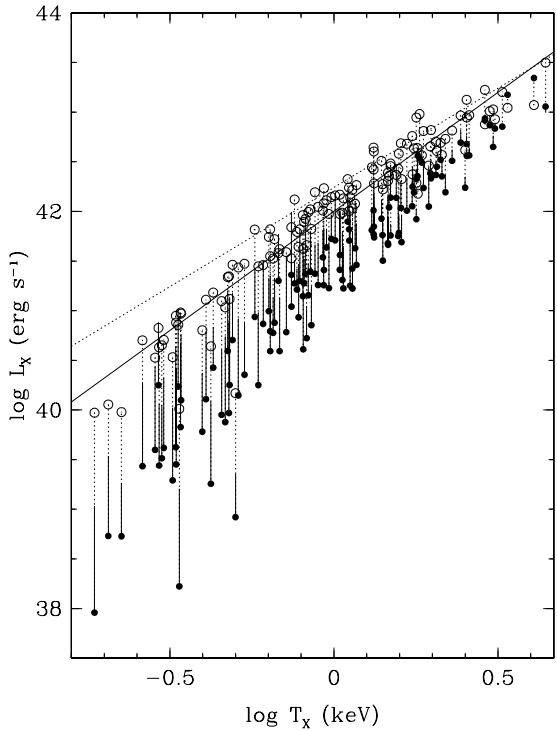}
\caption{$L_X-T_X$ relation showing the relative importance of variations
in mean hot gas density and clumping factor among our groups.  The solid
points are reproduced from Figure~\ref{fig: scaling} (but not labeled by
the number of member galaxies), while the open circles
include correction factors to force all groups to have the same mean hot
gas fraction and clumping factor (see eq.~\ref{eqn: lxtest}).  
Self-similar scaling is shown as the dotted line, and the best fit
to the open circles is shown as the solid line, having a slope of $2.4\pm 0.2$.
The two portions of the vertical line segments represent the separate
corrections due to hot gas fraction (solid) and clumping (dotted).
\label{fig: lxtest}}
\end{figure}

Figure~\ref{fig: lxtest} shows the relative impact of these
two effects on the $L_X-T_X$ relation for groups in our simulation.
We focus on $L_X-T_X$ here because its deviation from self-similarity
is more evident than that of $L_X-\sdm$, at least above the break
(cf. Figure~\ref{fig: scaling}).  The solid points (at the bottom of the
vertical lines) represent the groups reproduced from Figure~\ref{fig:
scaling}.  The open circles represent the group luminosities
``corrected'' to a common gas fraction and clumping factor,
\begin{equation}\label{eqn: lxtest}
L_{X,\rm corrected} = L_X (0.5/f_{\rm hot})^2 (C_{\rm 3~keV}/C_{\rm hot}),
\end{equation}
where $f_{\rm hot}$ is the fraction of hot baryons in each group,
$C_{\rm 3~keV}$ is the typical clumping factor for 3~keV groups,
and $C_{\rm hot}$ is the clumping factor for the specific group
(cf. Figure~\ref{fig: baryfrac}).  If all groups had a hot fraction
of 50\% and a clumping factor of $\approx 20$, this correction would
have no effect.  Instead, there
is an increasingly larger correction for smaller systems.  The solid
portion of each vertical line represents the correction factor
due to the mean hot fraction, and the dotted portion represents the
remaining factor due to clumping.  
Figure~\ref{fig: lxtest} shows that the two effects are of comparable
magnitude, and that the combination of the two brings the group luminosities
into much better agreement with
the self-similar relation (diagonal dotted line),
though there is still some tendency for the lower mass groups
to be underluminous.
The best-fit relation for the ``corrected'' luminosities
(which are not, of course, the physically predicted luminosities)
has a slope of $2.4\pm0.2$.
The remaining discrepancy with the self-similar $L_X-T_X$ relation,
and some of the remaining scatter, probably arise from the 
temperatures themselves, since smaller groups tend to be hotter
than the self-similar $T_X-\sigma$ relation predicts
(see Figure~\ref{fig: scaling}).

\begin{figure}
\plotone{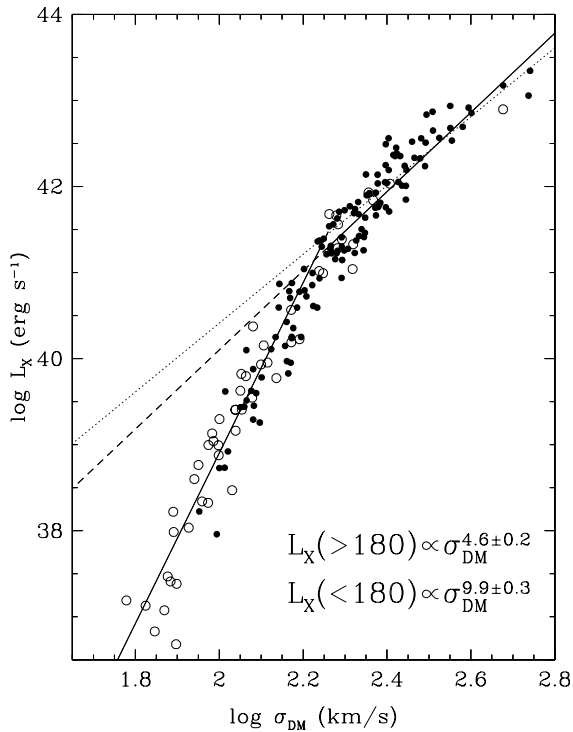}
\caption{Resolution test of the $L_X-\sdm$ relation, showing the groups
from the $(50\hmpc)^3$ simulation
(solid points, as in Figure~\ref{fig: scaling}
but not labeled by the number of member galaxies), and the higher
resolution, $(22.22\hmpc)^3$ simulation (open circles).  The agreement in
the region of overlap is very good, indicating that our results 
for the $L_X-\sdm$ relation are
not significantly affected by numerical resolution.  The scaling relations
of the combined sample are shown as the solid line, and listed in
the lower right.  The dotted line shows self-similar scaling normalized
to the largest systems.
\label{fig: restest}}
\end{figure}

A natural concern is that the clumping factor of our
smaller groups is artificially suppressed by resolution effects, since
our smallest systems contain as few as
$\sim 100$ hot particles.  Figure~\ref{fig: restest} shows a comparison
of $L_X-\sdm$ from our high-resolution simulation (open circles) to
the data shown in Figure~\ref{fig: scaling} from our large-volume simulation
(solid points).  We focus on $L_X-\sdm$ because $\sdm$ is a more
direct indicator of group mass than $T_X$.
While the mass range probed by the two simulations
is necessarily different, in the region of overlap they
agree remarkably well.  The best fit relation to the combined data is
shown in the lower right, and is in good agreement with the scalings
from Figure~\ref{fig: scaling}.  
If resolution were a major factor in the predicted departures from
self-similar scaling, we would expect low mass groups to be systematically
more luminous in the higher resolution simulation, and we would expect
the break at $\sdm \approx$180~km/s to shift down by a factor of two
to $\sdm \approx $90~km/s, corresponding to the
factor of eight increase in mass resolution.
While there is only one point well above the break from the high
resolution simulation, it sits close to the mean
$L_X-\sdm$ relation defined by the lower resolution, large-volume simulation.
The good agreement in Figure~\ref{fig: restest} demonstrates that the
reduced luminosity of low mass groups in our simulations is not
an artifact of finite resolution but arises instead from the
physical processes that the simulations incorporate.
This test also indicates that our two-phase decoupling
method for calculating luminosities is robust against modest changes
in resolution.

In summary, the breaks in the $L_X-\sdm$ and $L_X-T_X$ relations
are driven by the 
increased efficiency of radiative cooling in lower mass groups,
which affects both the fraction and the density structure 
of the hot gas.  The
fraction of $T>10^{5}$~K gas for our simulated groups drops from
50\% at $\sigma\approx 500$~km/s to 20\% at $\sigma\approx 100$~km/s,
which explains the majority of the decline in luminosity relative to
the self-similar prediction, while the remaining reduction is due to
smaller systems having lower clumping factors $C_{\rm hot}$.
This trend in hot fraction is consistent with observations, though the 
hot gas fractions of our large-volume simulation appear somewhat too low
once surface brightness biases are taken into account.
The hot gas fractions in our high-resolution simulation 
agree very well with observationally inferred values, and
the $L_X-\sdm$ relations of the two simulations agree very well
in their range of overlap.

The trend of a declining hot fraction in the group mass regime seems to
be a generic feature of CDM-based galaxy formation, arising in 
hydrodynamic simulations
of galaxy formation (\citealt{kay00}; \citealt{bla00}; \citealt{pea01}; this 
paper)
and semi-analytic models \citep{bow01}.
At the present epoch, the cooling time in the intragroup medium is
long, but during the early stages of their assembly, a larger fraction of
baryons were able to cool in smaller systems.  This rather straightforward
physical process just happens to become important in the mass regime of
groups, and it results in a breaking of self-similarity in the observed
scaling relations.

\section{Profiles}\label{sec: profiles}

Radial profiles of physical quantities such as temperature, surface
brightness, and electron density can provide insights into galaxy
formation processes within groups.  
In this section we present our
simulation results for these profiles, fit beta models to the surface
brightness profiles, and compare with observations, including the
entropy-temperature relation derived in PCN99.

\subsection{Average Profiles}

\begin{figure}
\plotone{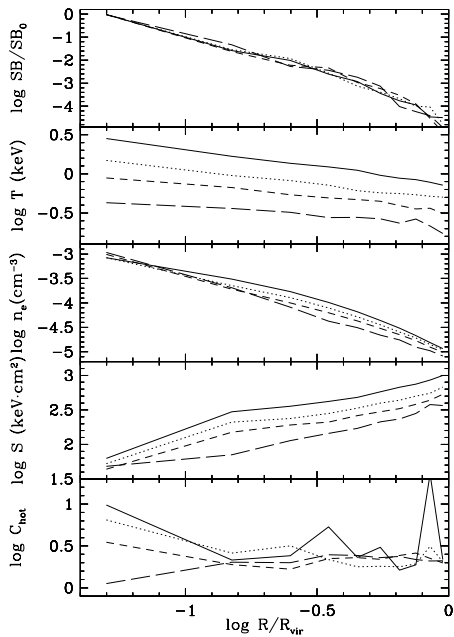}
\caption{Radial profiles of surface brightness (top panel),
temperature (second), electron density (third), entropy
(fourth), and clumping factor (bottom) for the simulated groups,
divided by mass
into four equal number subsamples (solid, dotted, short-dashed, and 
long-dashed lines, from highest to lowest mass).  The surface brightness and
temperature profiles are projected, averaged in cylindrical shells,
while the other profiles are 3-d, averaged in radial shells.
\label{fig: profiles}}
\end{figure}

Figure~\ref{fig: profiles} shows the average surface brightness,
temperature, electron density, entropy, and clumping factor profiles for
our simulated groups,
divided into four mass bins with 32 groups per bin.  Profiles
are scaled to the group Virial radius, then averaged.  The surface
brightness profiles (top panel) are shown scaled to the group central
surface brightness.  They are generally similar in form down to our
smallest groups.  This result contrasts with, e.g., Figure~1 of PCN99,
where the surface brightness profile is significantly shallower for
smaller systems.  However, MZ98 find the shapes of surface brightness
profiles for poor groups to be quite similar to those of clusters,
consistent with our simulated profiles.  We will examine these issues
in more detail when we discuss beta models in the next section.

Our temperature profiles (second panel) show a slight drop from the inner
to outer parts.  We do not produce a cool central region often seen in
groups (e.g., HP00) or cooling flow clusters \citep[e.g.,][]{all01};
we will return to this issue in \S\ref{sec: bspec}.  However,
beyond the innermost regions, both group and cluster temperature
profiles are observed to have a slight drop with radius \citep[MZ98;
HP00;][]{deg01}.  In clusters, where one is able to probe almost out to the
Virial radius, BeppoSAX observations by \citet{deg01}
indicate an isothermal core (for non-cooling flow clusters) out to $0.2
R_{\rm vir}$, and then a 30-40\% drop out to $0.5 R_{\rm vir}$, which
they claim is a steeper drop than predicted by numerical simulations.
From $0.2R_{\rm vir}$ to $0.5R_{\rm vir}$, our smallest groups show a
20\% drop in temperature, while our larger groups show a 40\% drop,
consistent with their observations.  Out to the Virial radius, the
drop is around a factor of two.  Thus, with minor discrepancies, our
simulations generally reproduce observed temperature profiles outside
of $0.2R_{\rm vir}$, predicting temperature profiles that are falling
slowly towards large radius.  We have presented projected temperature
profiles in Figure~\ref{fig: profiles} to allow more straightforward
comparison to observations, but our conclusions from the 3-D profiles
would not be significantly different.

The electron density profiles (middle panel) for the three subsamples
would lie on top of each other in a purely self-similar model.  The fact
that smaller groups have lower $n_e$ is a signature that cooling
has operated more efficiently in these systems, as we showed in
Figure~\ref{fig: baryfrac}.  Figure~\ref{fig: profiles} shows that cooling
preferentially removes hot gas at intermediate radii.  It is here that
the cooling time is comparable to a Hubble time, and hence there is
the most dramatic difference between large and small systems.  In the
central region, the cooling time has been short compared to a Hubble
time over much of the group's life, while in the outskirts it is much
longer than a Hubble time; thus in both regimes the electron density is
similar across all groups.  The shapes of our simulated density profiles
are markedly different from those predicted in a pre-heating scenario,
where the electron density profile would show an increasingly large core
in smaller systems.  Unfortunately there are no direct observations of
the electron density profile, only estimates from beta-model fitting
of surface brightness profiles (PCN99).  The combination of rapidly
improving Sunyaev-Zel'dovich and X-ray observations of groups may yield
electron density profiles in the future.

The entropy profile (second from bottom) is computed by taking the
mass-averaged value of $T/n_e^{2/3}$ of all particles within each bin.
The entropy profile declines smoothly from the outer edge into the center,
in agreement with simulations by \citet{tho01} and analytic expectations
based on convective equilibrium \citep{voi02}, though they differ from
those in the analytic model of \citet{bab01}.  We will discuss entropy
further in \S\ref{sec: entropy}.

The clumping factor profile (bottom panel) shows 
$\sum \rho_i/N_{\rm hot}\bar{\rho_{\rm hot}}$
(cf. eq.~\ref{eqn: clump}, but now the sum is over particles in
each radial bin, and $N_{\rm hot}$ and $\bar\rho_{\rm hot}$ are the
particle number and mean density of hot gas in the bin).
The largest systems
show a noisy profile in the outer parts, presumably because these objects
are dynamically younger and have more subclumps due to recent merging.
However, the systematic mass dependence of these profiles arises mainly
at small radii, where larger systems show significantly more clumping.
It is this inner region that drives the trend seen in the
bottom panel of Figure~\ref{fig: baryfrac}, and consequently (a
portion of) the luminosity drop in Figure~\ref{fig: lxtest}.
In part this trend arises because the largest systems often contain a
single dominant galaxy in the center, whereas in smaller systems there
can be significant offsets between the group center and the largest
galaxy's position.  
The lower $C_{\rm hot}$ values at small radii in smaller groups do not
appear to be resolution artifacts, since the luminosities predicted
by the high-resolution and large-volume simulations agree
(Figure~\ref{fig: restest}) and the electron density profiles
do not flatten in the centers of small groups, as would be expected
if resolution effects were important.

In summary, profiles may have greater power to discriminate between models
of group and cluster formation than scaling relations alone, since the 
latter can be adequately reproduced in both cooling and pre-heating models.
In particular, the surface brightness profile appears to be a key
diagnostic, with initial indications being that our simulations may be in
disagreement with observations.  We now examine this issue in more detail.

\subsection{Beta Models}\label{sec: bfit}

The canonical form for representing an X-ray surface brightness profile
is a ``beta model".  In this model, the parameter $\beta$ represents
the ratio of the specific energy in dark matter to that of the hot gas, namely
\begin{equation}\label{eqn: beta}
\beta = \frac{\mu m_p \sigma^2}{k_B T},
\end{equation}
where $\mu$ is the molecular weight and $m_p$ is the proton mass.
For an isothermal sphere in hydrostatic equilibrium with a King
model density profile, the resulting surface brightness profile is
\begin{equation}\label{eqn: sbprof}
S(r) = S(0) \bigl(1+r/r_{\rm core}\bigr)^{-3\beta+0.5},
\end{equation}
where $S(r)$ is the azimuthally averaged surface brightness at projected
radius $r$, and $r_{\rm core}$ is the core radius.  The electron density
distribution also follows a similar form, with an exponent of $-3\beta/2$.

Observationally, $\beta$ may be estimated either via equation~(\ref{eqn:
beta}), inferring the gas temperature from the X-ray spectrum
and obtaining the velocity dispersion or mass independently, or via
equation~(\ref{eqn: sbprof}), by fitting surface brightness profiles.
These spectroscopic and imaging estimates of $\beta$ are usually
referred to as $\bspec$ and $\bfit$, respectively.  Unfortunately,
the two estimates have historically yielded discrepant answers even
in well-studied clusters, with $\bfit\approx 2/3$ \citep{moh99}
and $\bspec\approx 1$ \citep[see discussion in][]{mul00}.  
Using hydrodynamic
simulations, \citet{nav95} find that fitting surface brightness profiles
yields $\bfit\approx 0.8$ even when $\bspec\approx 1$ because the
dark matter mass profiles deviate systematically from a King model.
Furthermore, they demonstrate that fitting to a smaller fraction of
the Virial radius (as often done in observational analyses, especially
of groups) biases $\bfit$ even lower.  These biases were invoked to
reconcile observations of $\bspec$ and $\bfit$.  
However,
recent {\it Chandra} observations of six large clusters with lensing
masses suggest $\bspec\approx 0.7$ \citep{all01}, so there may
be no discrepancy to be explained after all.

Determinations of $\bfit$ for groups are even more unsettled, primarily
because only a small portion of the surface brightness profile is
visible in these faint systems, and also because the contribution
from a central galaxy, which is excluded to obtain the intragroup gas
properties, is typically more important than in clusters.
HP00 find
a median $\bfit=0.46$ from 24 groups with $T_X\approx 0.4-1.6$~keV.
This implies considerably flatter surface brightness profiles than in
clusters, a fact that is the basis for the ``entropy floor" of PCN99.
However, MZ98 find a median $\bfit\approx 0.8$ for 9 groups, which is
{\it higher} than typical observed
values for clusters.  Both MZ98 and HP00 use two-component beta
models that are statistically warranted by the data, and in fact they
consider many of the same groups.  The fits of HP00 for groups in common
with MZ98 yield much smaller core radii and smaller $\bfit$, suggesting
a strong degeneracy between $r_{\rm core}$ and $\bfit$ \citep{mul00}.
HP00 claim to see a trend for lower $\bfit$ in smaller systems, but
this could be an artifact of the bias mentioned above, where fits to a
smaller fraction of the Virial radius yield lower $\bfit$ values, since
observations of their smallest groups extend to only $\sim 10-20\%$
of the Virial radius.  Thus, in our view, the case for flatter surface
brightness profiles in smaller groups has not been clearly proven.

\begin{figure}
\plotone{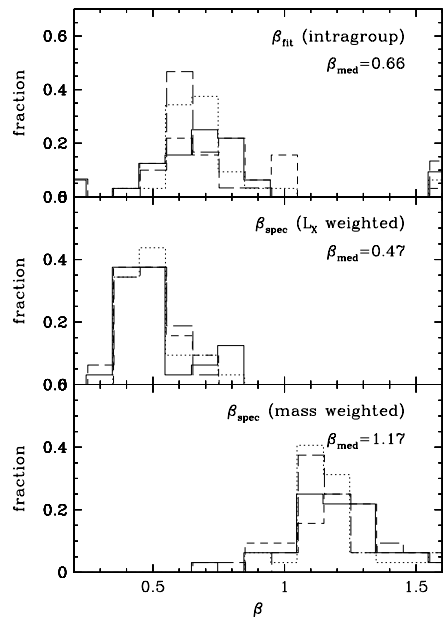}
\caption{{\it Top panel:} Histogram of $\bfit$ for the intragroup surface
brightness component of two-component beta models fit to our groups.
Median $\bfit$ value is indicated in upper right.  {\it Middle panel:}
Histogram of $\bspec$ for luminosity-weighted group temperatures. {\it
Bottom panel:} Histogram of $\bspec$ for mass-weighted average group
temperature.  In each panel, the four histograms correspond to four
mass bins containing nearly equal numbers of groups, with solid,
dotted, short-dashed, and long-dashed lines in decreasing order of mass.
\label{fig: beta}}
\end{figure}

Figure~\ref{fig: beta} shows a histogram of $\bfit$ values from
two-component beta model fits to our simulated group surface brightness
profiles.  The four line types correspond to our equal number subsamples.
All groups with $\bfit>1.6$ are stacked in the rightmost bin.  We allow
the two components to simultaneously vary freely in our Marquardt
minimization routine, then take the $\beta$ value associated with
the component having the larger $r_{\rm core}$, since it presumably
represents the intragroup component\footnote{In the case of groups,
where the central galaxy may make a significant contribution to the
gravitational potential, we are not convinced that it is sensible to fit
the ``galaxy'' as a physically distinct component, but here we follow
the standard practice of observational analyses.}. The median values of
$\bfit$ and $r_{\rm core}$ are 0.66 and $0.2 R_{\rm vir}$, respectively.
The second component (not shown) has median $\bfit\approx 0.5$ and
$r_{\rm core}\approx 0.03 R_{\rm vir}$.

Our median $\bfit=0.66$ is intermediate between that of HP00 and MZ98.
We find only a slight trend with size: our subset of largest groups has
a median $\bfit=0.68$, while our subset of smallest groups has median
$\bfit=0.62$.  We find that for many groups, the two-component beta model
is not a good fit even though sensible values of $\bfit$ are returned.
Since groups are not strictly isothermal, do not follow King model
profiles, and are typically not spherically symmetric (especially
smaller systems), perhaps this result is not surprising.  It may,
however, explain why the observed fits are so sensitive to the details of the
fitting procedures.  Further evidence that beta models are inappropriate
comes from a comparison of the surface brightness and electron density
profiles in Figure~\ref{fig: profiles}; though the former have a similar 
shape across all systems, the latter do not.

In principle, surface brightness profiles offer an excellent diagnostic
of intragroup gas physics.  Unfortunately, current observational
determinations of $\bfit$ appear too uncertain to validate or invalidate
any models.  This situation is likely to change with {\it XMM} and {\it
Chandra}, given their increase in sensitivity and spatial resolution.
If observations of flatter surface brightness profiles in smaller groups
persist (i.e., if $\bfit$ correlates with temperature), they would pose
a serious challenge for our simulation predictions.  Similar values of
$\bfit$ in groups and clusters, on the other hand, would strongly favor
the scenario presented here, in which cooling rather than pre-heating
is the primary cause of deviations from self-similar scaling.
Directly measured profiles will be more informative than model fits,
since beta models implicitly incorporate a number of assumptions
that may not be true in detail.

\subsection{Group Temperatures and $\bspec$}\label{sec: bspec}

The middle and bottom panels of Figure~\ref{fig: beta} show histograms
of $\bspec$ from our simulated groups.  The middle panel shows $\bspec$
when the usual luminosity-weighted temperature is used, and the bottom
panel shows the case when the mass-weighted temperature is used; in both
cases we have used $\sdm$ for the velocity dispersion.  Interestingly,
the luminosity-weighted case shows significant deviations from the
expected $\beta=1$, with a median value of $\bspec=0.47$.  Thus, the
luminosity-weighted temperatures are roughly twice the groups' Virial
temperatures.  Conversely, the mass-weighted temperatures underestimate
the Virial temperature by $\sim 20\%$.  There is no systematic trend of
$\bspec$ with group mass in either case.

Adiabatic simulations yield $\bspec\approx 1$ \citep[e.g.][]{nav95}.
As emphasized by \citet{all01}, this value appears high compared to the
latest {\it Chandra} observations of clusters, where $\bspec\approx
0.7$.  Hints of this
discrepancy were already present in earlier ROSAT data \citep[see][for
a summary]{tho01}.  In the simulations of \citet{tho01}, either cooling or
pre-heating produces increased temperatures compared to the adiabatic
case, yielding better agreement with the data.  
With pre-heating, the reason for the increased temperature is
obvious, but in the case of cooling, this nonintuitive result arises
because gas is cooled out of the center of the cluster, and the lowered
pressure support draws in hotter gas from the outskirts \citep{pea00}.
The cluster simulation with cooling by \citet{lew00} produced a similar
result.  In both cases, however, the derived value of $\bspec$ was
consistent with the latest observations, while our groups still appear
too hot.

We conclude that our predicted luminosity-weighted temperatures are
somewhat at odds with observations.  We investigated whether the procedure
of fitting Raymond-Smith models to coarsely binned X-ray spectra yielded
some systematic temperature bias, and found that when groups have a
temperature gradient towards the center (as in our larger systems),
the fitted temperature underestimates the true luminosity-weighted
temperature.  However, the magnitude of this effect is quite small
($\sim 10\%$), and in any case it cannot explain the discrepancy in our
smaller groups, which are nearly isothermal.

Instead, we believe the source of the discrepancy lies in the temperature
profiles of our simulated groups (cf. Figure~\ref{fig: profiles}).  In the
sample of HP00, about half the groups show a clear drop in temperature,
by as much as a factor of two, within $\sim 10-20\%$ of the Virial radius,
and the majority of remaining groups are at least consistent with having
a cooler central region.  Similar results are seen in large clusters,
though the drop in temperature is seen out to a smaller fraction of the
Virial radius \citep{all01}.  Our simulated groups, however, show no
such drop, and since the emission is dominated by the central region,
this leads to a much higher luminosity-weighted temperature.  If we
alter the temperatures of simulated group particles ``by hand'' so that
they are lower in the center by an amount consistent with observations,
then our median $\bspec$ increases to roughly 0.8, in agreement with
observed values.  (Specifically, we apply a linear temperature gradient
such that the central temperature is half that at $0.2 R_{\rm vir}$.)

It is not clear why our simulated groups do not have a cool central
region, while observed groups do.
If the physical process(es) that produce such a region are
analogous to those in clusters, then the insights gained from recent
cluster studies may provide clues.  For instance, recent observations
of cooling flow clusters suggest local isothermality in the central
region \citep{boe01a}, with little emission from the cold component
\citep{pete01}, in conflict with the standard cooling flow model
\citep{fab94}.  These findings suggest that some unidentified heat
source must counteract the cooling process, and serve to maintain the
central gas in a quasi-static, slightly cooler state.  This heat source
may be AGN in the central galaxy \citep{boe01b}, heat conduction from
the outer region \citep{nar01}, outflow induced shocks that can be
seen in recent Chandra images, or some as yet undiscovered process.
For instance, 
AGN are observed to produce ``bubbles'' in the intracluster medium, the walls of which
contain denser, cooler gas \citep{nul02}; any emission weighted measure of
temperature will be biased low by this gas since the emission depends on
the density squared.
If this or similar processes are operating in the centers of groups
(at a proportionally reduced amplitude that would make them difficult
to detect directly), then our simulations would be missing the input
physics required to properly model the central regions in these systems.
Alternatively,
we note that the outer radius of the cool region is roughly the radius
at which the metal-line cooling rate becomes smaller than a Hubble time
(while with only the primordial cooling assumed in our simulation, the
cooling time is still much longer).  Hence it is possible that including
metal-line cooling in our simulation would result in a cooler center,
particularly if there is a metallicity gradient towards the center.
However, as noted earlier, increased cooling does not necessarily
reduce gas temperatures, since reduced central pressures allow hotter
gas to flow in from the surrounding regions and experience adiabatic
heating along the way.

\subsection{Entropy}\label{sec: entropy}

Entropy is a powerful diagnostic of intragroup gas,
with a closer connection than temperature or density alone
to the processes that determine its physical state \citep{bow97,voi02}.
The fact that groups appear to show excess entropy relative to that
predicted purely from gravitational shock heating (PCN99) has been
used to argue for the presence of non-gravitational heating processes,
since radiative cooling reduces rather than increases gas entropy.
However, a more complete analysis suggests that cooling can effectively
raise the entropy inferred from X-ray data because the gas that
cools onto galaxies is no longer observable in the X-rays at all;
more efficient cooling in smaller systems can therefore produce an
apparent ``floor'' in group entropies \citep[C. Norman 2000, private comm.;][]{voi01}.

\begin{figure}
\plotone{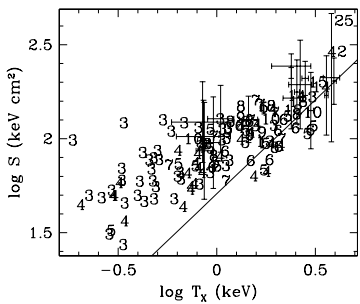}
\caption{ Entropy at $0.1R_{\rm vir}$ vs. X-ray temperature of simulated
groups.  The solid line shows the self-similar prediction, and the data
points are taken from PCN99.
\label{fig: ent}}
\end{figure}

Figure~\ref{fig: ent} plots the entropy at 10\% of the Virial radius
against group X-ray temperature, for the simulated groups and the 
observational analysis of PCN99.  
The group entropy profiles (Figure~\ref{fig: profiles}) are reasonably
well described by power laws.  We compute entropies for Figure~\ref{fig: ent}
by fitting each group's profile with a power law, then taking the value 
of the fit at $R=0.1R_{\rm vir}$.  The solid line shows the self-similar 
entropy relation derived using adiabatic simulations (see PCN99), $S(0.1R_{\rm
vir})=45(T/{\rm keV})(f_{\rm gas}/0.06)^{-2/3}h^{-4/3}\entunit$, with $f_{\rm
gas}=\Omega_b/\Omega_m=0.118$ in our case.

It is clear that the simulated group entropies
do not follow the 
self-similar relation, and are instead in reasonable agreement with
the data (given their large error bars).  This agreement is somewhat
surprising given that our surface brightness profiles do not 
show the strong mass dependence inferred by PCN99, and it once again suggests
that the canonical beta model may not be an appropriate description
of poor groups.  Our results are also in agreement with the analytic model of
\citet{bry00}, which is based on a declining hot fraction
with group mass similar to that predicted by our simulation.  Thus, it
appears that the physical mechanism of cooling out low-entropy gas and
leaving an effective ``entropy floor" \citep{voi02} is operating in
our simulations.  Note, however, that the simulation does not indicate a
hard floor at $\sim 100$~keV~cm$^2$, but rather a 
decline of entropy with group temperature that is slower
than predicted by self-similar models.

\section{Modeling Observed Systems}\label{sec: obseff}

In the real universe, it is not possible to directly observe the
luminosity, temperature, and dark matter velocity dispersion of a
zero-metallicity intragroup medium out to the Virial radius.  Hence,
to compare theoretical predictions to observations, we must model
observational effects in our simulations.  In this section we consider
some ``real-world" effects and assess how they affect the inferred
scaling relations and derived interpretations.  Our analysis here
is simple and intended only to indicate the sign and rough
magnitude of such effects; we defer a detailed side-by-side comparison of
simulated and observed groups to future work.  We will see that there
are non-trivial differences between the scaling relations of ``observed"
groups and the idealized ones discussed in \S\ref{sec: scale}, but that
none of our fundamental conclusions are altered as a result.

\subsection{Metallicity}\label{sec: metals}

\begin{figure}
\plotone{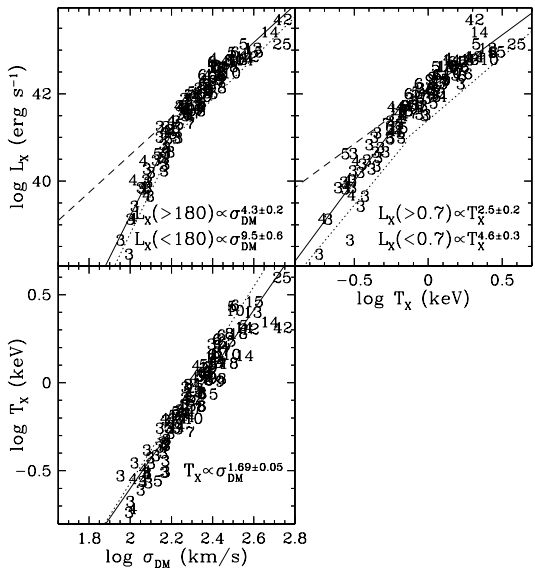}
\caption{ Scaling relations as in Figure~\ref{fig: scaling}, but the X-ray
luminosities and temperatures have been computed using a metallicity
given by equation~(\ref{eqn: metal}).  The best-fit relations from 
Figure~\ref{fig: scaling} have been reproduced here as the dotted lines.
\label{fig: metal}}
\end{figure}

Since metal lines provide a dominant contribution to the total luminosity
at temperatures below a few~keV, they could in principle significantly
affect the predicted luminosity scaling relations.  We include metals
via a fit to the metallicity vs. temperature data in Figure~1 of
\citet{davis99}, namely
\begin{equation}\label{eqn: metal}
\log{Z} = 1.04 \log{T_X} -0.73,
\end{equation}
where $Z$ is the metallicity in solar units, and $T_X$ is in keV.
We cap the metallicity at the cluster value of 0.3 solar; systems above
1.6~keV are thus set to this metallicity.  Note that \citet{davis99} derive
a much steeper dependence of abundance on temperature for systems with
$T_X<1.5$~keV ($Z\propto T^{2.5}$).  However this relation would produce
unrealistically low metallicities for our smallest groups, so instead we
use our fit to their entire data set up to $T_X\approx 3$~keV.
While there has been some controversy regarding metallicities derived
from {\it ASCA} data \citep[see, e.g.,][]{buo00}, the low metallicities
in the \citet{davis99} sample have now been confirmed in several cases
with improved data from {\it XMM} (Mulchaey, private comm.).  Thus,
the drop in metallicity below 2~keV appears to be real, though it is
not straightforward to explain \citep{mul00}.

Figure~\ref{fig: metal} shows our predicted scaling relations after
incorporating equation~(\ref{eqn: metal}) into our calculations
with the Raymond-Smith code.  The relations continue to show the same
qualitative behavior, and the break values are not changed significantly.
The luminosity relations are shallower above the break, since the
metallicity is mostly constant in these larger systems, and metal lines
provide a greater contribution in smaller (cooler) groups.  Below
the break, the relations are steepened because the metallicity is dropping
with group mass.  At 1~keV, the luminosity is higher by a factor $\sim
3$ compared to the zero-metallicity case (compare the dotted and solid
lines in the luminosity relations), indicating the dominant contribution
of line emission.  The $T_X-\sdm$ relation is also slightly affected,
as higher mass systems have lower luminosity-weighted temperatures; this
is because the relative luminosity weighting of low-temperature gas in
the system increases with the presence of metals in this temperature
regime.  However, these small differences do not qualitatively affect
our conclusions drawn from the zero-metallicity case (Figure~\ref{fig:
scaling}).

While it appears that the metallicity in groups is lower than in clusters,
this difference is not well established.  We therefore test another
model using a constant metallicity of $0.3Z_\odot$ in all groups, and we
list the resulting scalings in Table~1.  This calculation shows that if
the metallicity is similar in groups and clusters, then the $L_X-\sdm$
and $L_X-T_X$ slopes are shallower in all regimes.  Even in this case,
however, the breaks in the luminosity scalings remain quite evident.

\subsection{Surface Brightness Effects}\label{sec: extent}

\begin{figure}
\plotone{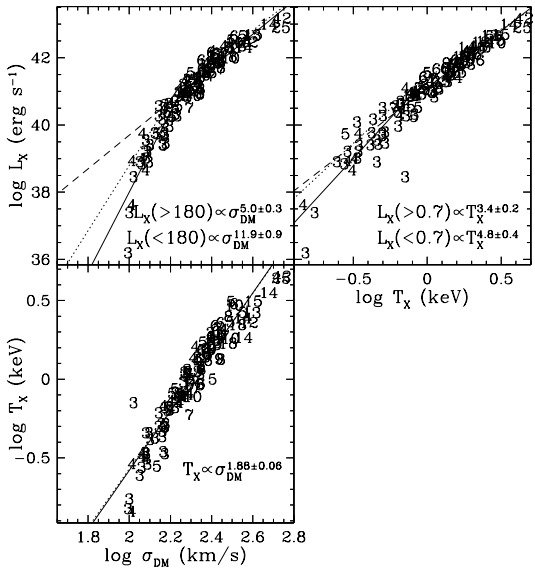}
\caption{Scaling relations as in Figure~\ref{fig: scaling}, but the X-ray
luminosities and temperatures have been computed only out to an
``observable" fraction of the Virial radius as given by
equation~(\ref{eqn: rx}).  These calculations assume zero metallicity.  
The best-fit relations from 
Figure~\ref{fig: scaling} have been reproduced here as the dotted lines.
\label{fig: sb}}
\end{figure}

While current cluster observations can reach the Virial radius and
sometimes beyond, this is rare in groups \citep{mul00}.  Thus, group
observations only probe the innermost regions, and it is not straightforward
to compare their properties with clusters where physical quantities
are measured over the entire system.  We investigated how this difference
affects derived baryon fractions in \S\ref{sec: baryfrac}, and we now consider
its effect on scaling relations.

Figure~\ref{fig: sb} shows the result of applying the surface brightness
cut in equation~(\ref{eqn: rx}) to our groups.  Note that we have
reset the metallicity to zero to isolate the effects.  Comparing to
Figure~\ref{fig: scaling} (whose best-fit scalings are reproduced as
the dotted lines), the luminosity relations are somewhat steeper, as is
expected since the surface brightness cut is stronger in smaller systems.
But the effect is fairly mild, even in small groups where the fraction
of the Virial radius probed is 20\% or less.  The $T_X-\sdm$ relation
is essentially unaffected.

The reason that the surface brightness cut has only a mild effect
is that most of the group luminosity comes from the central region
(cf. Figure~\ref{fig: profiles}), so there is minimal impact until the
cutoff radius becomes a fairly small fraction of the Virial radius.
Still, one must be cautious about interpreting observations where less
than 20\% of the group's Virial radius is being probed, especially because
the scatter in X-ray surface brightness between groups will select out
the brightest groups of a given mass.  Note that we did not include 
scatter in the X-ray extent relation used in equation~(\ref{eqn: rx}),
whereas observations show a sizeable scatter \citep{mul00}.  Overall,
however, current observations are sensitive enough to probe scaling
relations down to fairly small groups reasonably reliably, assuming that
issues of point-source contamination and Galactic foreground 
can be overcome.

\subsection{Velocity Dispersion Bias}\label{sec: sigma}

In computing $L_X-\sigma$ and $T_X-\sigma$ relations in \S\ref{sec:
scale}, we have used the dark matter velocity dispersion $\sdm$.
However, the observed velocity dispersion is typically computed from galaxies,
which may be biased relative to the dark matter, and this can influence
the interpretation of the observed correlations.  In this section we
examine this bias and its effect on the observed scaling relations.

\begin{figure}
\plotone{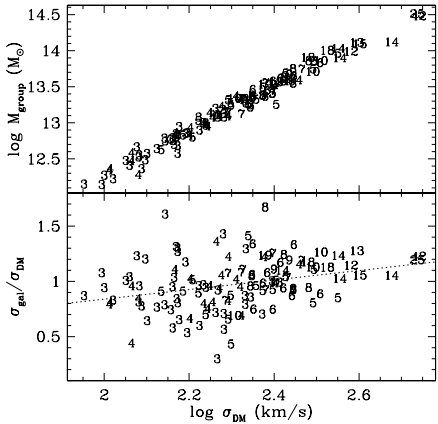}
\caption{{\it Top panel:} Group mass vs. dark matter velocity dispersion.
{\it Bottom panel:} Ratio of velocity dispersion calculated from
galaxies to that of the dark matter.  The dotted line shows a linear
fit to the data points.
\label{fig: velbias}}
\end{figure}

The top panel of Figure~\ref{fig: velbias} shows the group mass
vs. $\sdm$.  There is a very tight correlation, as would be expected
for relaxed systems.  As before, the plot symbols indicate the number of
galaxies in each group.  The bottom panel shows the ratio of the galaxy
velocity dispersion $\sgal$ to $\sdm$, where
\begin{equation}\label{eqn: sgal}
\sgal^2 = \frac{1}{N_{\rm gal}-1}\sum^{N_{\rm gal}}_{i=1} |
{\bf v}_{{\rm gal},i}-\langle {\bf v}_{\rm gal}\rangle|^2 ~,
\end{equation}
and $\langle {\bf v}_{\rm gal}\rangle$
is the (unweighted) mean velocity of the galaxies.
There is a weak but discernible trend for smaller systems to have their
velocity dispersion underestimated.  The dotted line shows the best fit
linear relation, which suggests that on average, at $\sdm=100$~km/s, the
galaxy velocity dispersion underestimates the true velocity dispersion
by 15\%.  Conversely, the galaxy velocity dispersion of large groups
is slightly higher than $\sdm$ on average.  Moreover, there is an
increasing scatter to smaller systems, since there are only a few
galaxy tracers with which to estimate the dispersion.
This scatter could be larger still in observational studies because of
projection effects \citep{tov02}.  
Hence, using galaxy velocity dispersions, particularly in systems with few
identified members, may have an impact on the velocity dispersion scaling 
relations.

%

\begin{figure}
\plotone{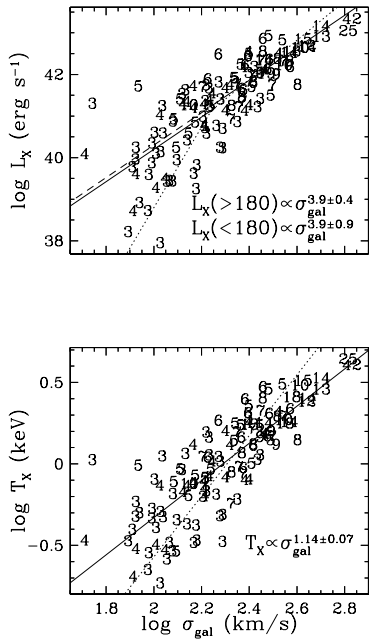}
\caption{Velocity dispersion scaling relations
with the velocity dispersion computed from the galaxies rather than the
dark matter.  The best-fit relations from Figure~\ref{fig: scaling} are
shown as the dotted lines.
\label{fig: sgal}}
\end{figure}

Figure~\ref{fig: sgal} shows the simulated group
scaling relations $L_X-\sgal$ and $T_X-\sgal$.  $L_X-T_X$ is not shown
since it is (obviously) unchanged from Figure~\ref{fig: scaling}.
The velocity dispersion scalings, on the other hand,
are substantially altered.  $L_X-\sgal$ shows
a somewhat shallower slope consistent with self-similarity, and the 
fitted slope
below the previous break value of 180~km/s is now identical to
the slope above it. Meanwhile,
$T_X-\sgal$ shows a relation much shallower than obtained using $\sdm$,
with $T_X\propto\sgal^{1.13}$.  

\begin{figure}
\plotone{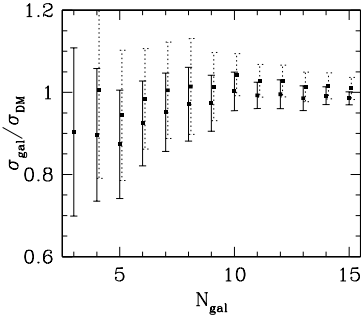}
\caption{Velocity dispersion calculated from the $N_{\rm gal}$
most massive galaxies within a group, divided by the velocity dispersion
of the dark matter.
Points with solid error bars
show the calculation using equation~(\ref{eqn: sgal}), while points with
dotted error bars indicate the effect of excluding the most massive
galaxy before computing $\sgal$.
Results are computed only for the 11 simulated groups that have ten or
more member galaxies.
The error bars represent the $1-\sigma$ scatter of $\sgal$
values within each bin; the error on the mean
of $\sgal/\sdm$ is smaller (e.g., for points at $N_{\rm gal}\leq 10$, it is
lower by a factor of $\sqrt{11}$).
\label{fig: sigmagal}}
\end{figure}

To examine this result in greater detail, we selected the eleven simulated
groups that have ten or more member galaxies, then took a subset of the
$N_{\rm gal}$ most massive galaxies in each group and recomputed $\sgal$.
In Figure~\ref{fig: sigmagal}, points with solid error bars show the
mean and the 1-$\sigma$ scatter of the ratio $\sgal/\sdm$ as a function
of the number of galaxies $N_{\rm gal}$ used in the $\sgal$ calculation.
At high $N_{\rm gal}$, the ratio is close to unity and the scatter is
small, but for low $N_{\rm gal}$ the scatter is large and the mean is
biased low.  If we
exclude the most massive galaxy from the velocity dispersion calculation
(points with dotted error bars), then the mean ratio remains close to
unity even at low $N_{\rm gal}$, which demonstrates that the bias in
$\sgal$ arises mainly from the tendency of the most massive, central
galaxy to move slowly with respect to the group center-of-mass velocity.
However,
if we recalculate the $T_X-\sgal$ relation using velocity dispersions
that exclude the most massive galaxy, we still get a fitted slope of
$\approx 1.1$.  This demonstrates that it is the {\it random scatter} in
the galaxy velocity dispersion estimates that affects the $T_X-\sgal$ slope,
not the small {\it systematic} bias of $\sgal$ with respect to $\sdm$.
Our fits to $L_X-\sgal$ and $T_X-\sgal$ are performed in log space
without accounting for random errors in the estimation of the true $\sigma$
from $\sgal$.  Since downward fluctuations reduce $\log\sigma$ more
than upward fluctuations increase it, the effect of these random errors
is to flatten the $L_X-\sigma$ and
$T_X-\sigma$ relations, especially in the low mass regime where the errors are 
larger.

Noisy velocity dispersion estimates may explain some conflicting
observational results regarding the $T_X-\sigma$ relation.  HP00 find
$T_X\propto \sgal^{0.9}$, and they interpret the higher-than-expected
temperatures at low $\sgal$ as evidence for pre-heating.  However, the
velocity dispersions of their smallest groups were typically computed
from 3--5 member galaxies (compiled from the literature).  It is not
strictly fair to compare their slope with our $T_X-\sgal$ slope, since
our galaxy samples have different ``completeness limits", but it is
illustrative that the slopes are similar.  Conversely, the MZ98 sample
includes only groups with over 20 spectroscopic members, and they find
$T_X\propto \sgal^{2.2\pm 0.9}$, in agreement with cluster samples.

In the currently relevant observational regime ($\sigma\ga 200$~km/s),
the $L_X-\sigma$ relation is not significantly affected by the scatter
in $\sgal$ (Figure~\ref{fig: sgal}, top).  HP00 find $L_X\propto
\sgal^{4.4}$, consistent with self-similarity, but they attribute this
to coincidentally canceling reductions in $L_X$ owing to pre-heating,
and in $\sgal$ owing to a noisy estimation from a small number of galaxies.
Our analysis shows that, at least for groups of this size, the $L_X-\sgal$
relation is unaffected within current observational errors.  In support
of this idea, MZ98 and \cite{zim01} also find $L_X\propto \sgal^{4.3-4.4}$
using many more galaxies per group.  \citet{tov02} argue that the velocity
dispersion bias arises because of geometrical projection effects in
1-D dispersion estimates, and they also find $L_X\propto \sgal^{4}$
when properly accounting for such effects.

In summary, velocity dispersion bias can have a significant effect on the
$T_X-\sigma$ relation, but only a mild effect on $L_X-\sigma$, at least
above the break.  Observations that have sufficient numbers of galaxies
per group ($\ga 10$) indicate little evidence for the anomalously
hot intragroup gas in poor groups that would be expected in a pre-heating
scenario.

\subsection{Comparison to Observations}\label{sec: obs}

Though there are still many fundamental differences between the way
our simulations and observations are analyzed, as well as some inherent
modeling difficulties in our simulation, it is nevertheless instructive
to compare our scaling relation predictions with available observations.
To model observations as closely as we can given current constraints,
we apply the metallicity relation as in equation~(\ref{eqn: metal}),
we include surface brightness effects as in equation~(\ref{eqn: rx}),
and we use the dark matter velocity dispersion.  The last choice is
motivated by the fact that the latest group samples have deep imaging
that permits identification of 10-50 group members \citep[e.g.][]{zim01}, which
should be sufficient to trace the true mass of the group much like $\sdm$,
though still with somewhat larger scatter.

\begin{figure}
\plotone{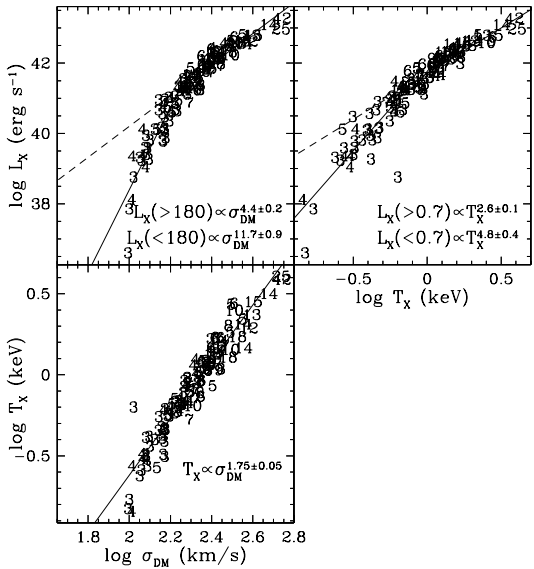}
\caption{Scaling relations as in Figure~\ref{fig: scaling}, but the X-ray
luminosities and temperatures have been computed using a metallicity
given by equation~(\ref{eqn: metal}) and a surface brightness
cut given by equation~(\ref{eqn: rx}).  This represents our attempt
to most closely match current observations, which are discussed
in the text.
\label{fig: scalingobs}}
\end{figure}

Figure~\ref{fig: scalingobs} shows the resulting scaling relations,
and the fits are also listed in Table~1.  The $L_X-\sdm$ relation slopes
above and below the break are close to those in a compilation of clusters,
groups and galaxies by \citet{mah01}, who found $L_X\propto \sigma^{4.4}$
down to $\sigma=350$~km/s and $L_X\propto \sigma^{10}$ below that down
to galaxy scales.  Our break occurs at a lower $\sigma$ than theirs,
but the observed 
relation is poorly constrained in the break region, and the data
sample is inhomogeneous, so the discrepancy may not be too serious.

The slope of the $L_X-T_X$ relation above the break is in good agreement
with MZ98 and \citet{zim01}, and it is also consistent with the observed
relation in clusters \citep{whi97}.  However, it is in poor agreement with
samples by HP00 and \cite{xue00}, who find $L_X\propto T_X^{4.9}$ and
$L_X\propto T_X^{5.6}$.  Those results are in better agreement with the
slope {\it below} our break at 0.8~keV.  These samples have $T_X\approx
0.4-1.6$~keV, thus their steep slopes suggest a higher break temperature,
perhaps $\sim 1-1.5$~keV.  As with our $L_X-\sigma$ relation,
our break appears to occur at a mass that is slightly too low.

Our $T_X-\sigma$ relation shows some deviation from self-similarity
($T_X\propto\sigma^{1.75}$), although as we saw in \S\ref{sec: metals},
this deviation arises mostly from metal line emission weighted into the
temperature determination, not from any physical process that actually
raises the gas temperature.  Metal line cooling is another way in which the
$T_X-\sigma$ relation can mimic excess temperature in $T_X\la 1$~keV
groups, in addition to the effects of scatter in $\sigma$ estimates from
small numbers of galaxies (\S\ref{sec: sigma}).  Observations suggest
that the $T_X-\sigma$ relation is fairly close to self-similar from
clusters down to groups \citep{mul00}, in agreement with our predictions.

So far we have not discussed the amplitudes of the various scaling
relations, only the slopes.  Our $L_X-\sigma$ relations amplitude is in
very good agreement with observations.  MZ98, HP00, and \citet{zim01}
all find $L_X= 10^{42.5}-10^{42.7}\lxunit$ at $\sigma=300$~km/s
where the observations are relatively reliable, while we obtain $L_X=
10^{42.5}\lxunit$.  This is encouraging, given that this (large-volume)
simulation cools an excessive fraction of baryons (see Figure~\ref{fig:
baryfrac} and the accompanying discussion), and we have made large
corrections to the luminosities based upon our two-phase decoupling (see
\S\ref{sec: sim}).  If our predictions were highly sensitive to these
effects, it would be quite a coincidence that they conspired to bring
our luminosity-mass relation into such good agreement with observations.

Conversely, the amplitudes of the temperature relations are in poor
agreement with observations.  As discussed in \S\ref{sec: bspec}, our
luminosity-weighted temperatures are higher than observed by a factor
of $\approx 1.5-2$ at a given $\sigma$.  The lower factor comes from
considering $\bspec$ as observed in {\it Chandra} data of large clusters
by \citet{all01} (with the caveat that our systems are much smaller than
any observed by those authors); the higher factor comes from comparing
the amplitude of our $T_X-\sigma$ relation with {\it ROSAT} observations.
For example, we predict a temperature of 1.7~keV at
$\sigma=300$~km/s, while
MZ98 and HP00 find 0.8 and 0.9~keV, respectively, for such systems.
It remains to be seen whether {\it Chandra} observations of these groups
will yield higher temperatures as has happened for some clusters.

In the $L_X-T_X$ relation, this overestimation of temperatures translates
to our predicted luminosities being too low by a factor $\sim 3-8$
at fixed $T_X$ (for $L_X \propto T_X^3$).  For example, at $T_X=1$~keV,
MZ98 find $L_X=10^{42.4}\lxunit$, HP00 find $L_X=10^{43}\lxunit$, and the
extrapolation of the \citet{whi97} relation yields $L_X=10^{42.7}\lxunit$;
our simulated groups yield $L_X=10^{41.7}\lxunit$ at this temperature.
As discussed in \S\ref{sec: bspec}, the {\it ad hoc} correction of
lowering the temperature of gas within $0.2 R_{\rm vir}$ in accordance
with temperature profiles seen by HP00 results in lowering our groups'
$T_X$ by $\sim$70\%, and such a reduction (regardless of what 
physical phenomenon is responsible for it) 
brings both our $T_X-\sigma$ and $L_X-T_X$ relations into
the observed range.

In summary, our simulation seems to do a good job of reproducing the
slopes of observed scaling relations, but the breaks in the $L_X-\sigma$
and $L_X-T_X$ relations appear to occur at a group mass that is somewhat
too small.  This discrepancy might be resolved by improvements in the
numerics or astrophysical modeling, or it could indicate that some degree
of pre-heating is still required, though much less than in a model with
adiabatic gas dynamics.  The amplitude of our
$L_X-\sigma$ relation agrees reasonably well
with observations, suggesting that our groups' luminosities 
are being accurately modeled.  However, our simulated X-ray temperatures
appear to be discrepant with observations, being too high by a factor of
$1.5-2$ at a given mass.  This discrepancy manifests itself
in both the $L_X-T_X$
relation and the $T_X-\sigma$ relation.  The lower temperatures of
observed groups appear to be connected to cooler central regions that
are not reproduced by our simulation.  We suspect that the cause of
this discrepancy is astrophysical, perhaps our use of zero-metallicity
cooling curves during dynamical evolution, or perhaps the absence of a
significant physical process such as AGN heating or conduction.

\section{Conclusions}\label{sec: concl}

Using a large-scale hydrodynamic cosmological simulation that is able to
resolve sub-$L_\star$ galaxies, we have investigated the X-ray
scaling relations for galaxy groups with velocity dispersions ranging
from $100-550$~km/s.  In particular, our goal was to assess whether these
scaling relations are consistent with those predicted by simple self-similar
models and adiabatic simulations.  Our main findings are:

\begin{enumerate}

\item Our luminosity scaling relations $L_X-T_X$ and $L_X-\sigma$
do not follow the self-similar form at any point in our mass range.
The simulated relations are always steeper than the self-similar
predictions, with lower luminosities at smaller masses.

\item These scaling relations show a ``break" around $180$~km/s
($0.7$~keV), below which the relations steepen even further.

\item The $T_X-\sigma$ relation shows no break, and its slope is
reasonably consistent with the self-similar model, indicating that these
groups are Virialized and in hydrostatic equilibrium.

\item In our simulations, self-similarity is broken mainly as a result of
the hot gas fraction falling with group size, from $\approx 50$\% of the
baryonic mass at $\sigma=500$~km/s to $\approx 20$\% at $\sigma=100$~km/s.
There is a secondary effect from the fact that the hot gas clumping
factor (mostly in the groups' central regions) drops by a factor of
roughly three over this mass range.

\item Our large-volume simulation cools too many baryons, resulting
in a typical hot fraction that is lower than observed by $\sim 40\%$,
once X-ray surface brightness thresholds are taken into account.  Our
high-resolution simulation predicts {\it more} hot gas at a given group
mass, and is in good agreement with observations.  Thus, the amplitude
of the $\fhot-\sigma$ relation is somewhat sensitive to resolution,
though not in the sense that one might naively expect.  The {\it trend}
of falling hot gas fraction towards lower group mass, which arises in
both simulations and accounts for most of the departure from self-similar
scaling relations, is consistent with available observations.

\item Our two simulations, which differ in mass resolution by a factor
of eight, predict very similar X-ray luminosities at fixed group mass.
To the extent that we can test them with these simulations, the
departures of the $L_X-\sigma$ and $T_X-\sigma$ relations from self-similar
scaling are not artifacts of numerical resolution: the reduced luminosities
of low mass groups are a result of physical processes that are modeled
consistently between the two simulations.

\item Surface brightness profiles offer an excellent diagnostic for
constraining gas physics in groups, in particular for distinguishing
cooling from pre-heating as the primary cause of departure from
self-similar scalings, but current observations are too uncertain to give
clear guidance.  Our groups' median $\bfit=0.64$ is intermediate between 
various
observations, and we predict no significant trend of $\bfit$ with group
size.  Our simulations do reproduce the observed entropy-temperature
relation (PCN99), though we do not predict a hard floor but merely
a shallower decline of entropy with group mass than predicted in
self-similar models.

\item Accounting for the effects of metallicity or for observational
biases due to X-ray surface brightness thresholds does not qualitatively
alter the conclusions listed above.

\item Noisy estimates of group velocity dispersions have a significant
effect on the derived $T_X-\sigma$ relation.  The most massive galaxy in a
group tends to move close to the group center-of-mass velocity, so its
inclusion in a velocity dispersion estimate biases $\sigma$ low when
the number of galaxies is small.  Even more important
is the scatter in $\sigma$ when the number of tracers is small,
which biases a simple fit of $T_X-\sigma$.
Our simulations suggest that $\ga 10$~galaxies are needed
to obtain velocity dispersions that are precise enough for $T_X-\sigma$
determination.
Noisy $\sigma$ estimates may account for some of the discrepancies between
apparently conflicting observational analyses.
The $L_X-\sigma$ relation can also be biased by noisy $\sigma$ estimates,
but this effect does not seem to be important in the mass range probed
by current data.

\item Accounting for observational biases, our scaling relation slopes
are roughly in agreement with observations, though our break appears at
somewhat too low a group mass.  With our best attempts to mimic current
observational approaches, our simulation predicts the following scaling
relations for galaxy groups with $\sigma\la 500$~km/s:

\begin{itemize}
\item $L_X\propto \sigma^{4.4}$ for $\sigma>180$~km/s, and \\$L_X\propto 
\sigma^{11.7}$ for $\sigma<180$~km/s.
\item $L_X\propto T_X^{2.6}$ for $T_X>0.7$~keV, and \\$L_X\propto T_X^{4.8}$ 
for $T_X<0.7$~keV.
\item $T_X\propto \sigma^{1.75}$ when $\ga 10$ galaxies used to estimate
$\sigma$, and\\$T_X\propto \sigma^{\approx 1}$ when 3-5 galaxies are used
for the smallest groups.
\end{itemize}

The relations with various assumptions about metallicity, surface brightness
thresholds, and velocity dispersion estimates are summarized in Table~1.

\item While the amplitude of our $L_X-\sigma$ relation amplitude is in
good agreement with observations, the amplitudes of our $T_X-\sigma$
and $L_X-T_X$ relations show significant discrepancies, and our
median value of $\bspec=0.47$ is lower than observational estimates.
We conclude that the luminosity-weighted temperatures 
of our simulated groups are too high by a factor of $1.5-2$.
This discrepancy appears to originate in the central regions,
with $r<0.2R_{\rm vir}$, where observed group temperature profiles
drop towards the center while our 
simulated profiles are flat or rising.  An {\it ad hoc} introduction of
a cool central region brings the amplitudes of both the $T_X-\sigma$
and $L_X-T_X$ relations into reasonable agreement with observations.
We suspect that the ultimate origin of these discrepancies is missing 
physics in our simulation, such as AGN heating of the intracluster gas,
but we cannot rule out numerical effects.

\end{enumerate}

These results can be taken on several levels.  At their most basic
and most robust level, they indicate that groups cannot be treated as
self-similarly scaled-down versions of clusters, since radiative cooling
plays an increasingly important role in smaller systems.  Although the
cooling time in the outskirts of groups and clusters is longer than
the Hubble time today, cooling times were shorter in the subsystems
that merged to make up the final group, and even today cooling can be
significant out to a non-trivial fraction of the Virial radius.  Thus,
the conventional approach of modeling X-ray properties with purely
adiabatic physics must be applied with caution, especially as one moves
from the mass regime of rich clusters to that of poor clusters and
groups.  Since cooling is a process that is known to occur, more exotic
processes like pre-heating or entropy injection should be examined in
the context of models that already incorporate cooling.  Unfortunately,
this requirement increases the complexity and the uncertainty of the
calculations, whether they are analytic or numerical.

At a second level, our results show that this simulation with cooling {\it
qualitatively} reproduces the trends seen in observed scaling relations,
namely a reduced luminosity of lower mass systems (relative
to self-similar predictions), and a break below which the luminosity
relations steepen further.  While we have used numerical methods to
obtain this result, it may be explained by a simple physical picture
based on hierarchical growth of structure.  At any epoch, the progenitor
of a group will have a lower Virial temperature than the progenitor
of a cluster.  Hence, integrated over the formation history of these
objects, a larger fraction of baryons will cool in the group environment than 
in the
cluster environment.  This results in the trends seen in the simulations
and in the observations.  Put another way, it is well known that in
clusters most baryons are hot, while in galaxies most baryons are cold;
our results simply suggest that the transition between these regimes
occurs in galaxy groups around 1~keV.

At a third, more uncertain level, our results may be taken as evidence
that a $\Lambda$CDM cosmological model with standard gas dynamics and
radiative cooling can {\it quantitatively} explain observed group scaling
relations, without strong pre-heating or substantial entropy injection.
The uncertainty arises because our simulation does not reproduce the
observations in their entirety: the forms of the predicted scalings
are about right, but the mass scale of the breaks is somewhat too low,
the cold gas fractions are somewhat resolution-dependent, and at a given
$\sigma$ the luminosity-weighted temperatures are too high.  Presently,
we do not know the extent to which these discrepancies reflect numerical
inaccuracies, missing astrophysical processes, incorrect cosmological
parameters, or errors in the observational inferences themselves.
We therefore do not know whether solving these problems will reduce the
role of cooling to the point that it no longer explains the observed form
of the scaling relations.  Definitive measurements of resolved surface
brightness and gas density profiles would be a powerful diagnostic
for the relative importance of radiative cooling and non-gravitational
heating in accounting for observed scaling relations.

Our simulation demonstrates that cooling can reproduce many of the
qualitative features in scaling relations that are often quoted as
evidence for pre-heating.  Our conclusion on this point agrees with
those drawn by \cite{bry00} and \cite{voi01} on the basis of analytic
models and by \cite{mua01} on the basis of numerical simulations.
New in this paper are the use of relatively high-resolution simulations
well suited to the group mass regime and a more detailed consideration
of observational issues than these earlier studies.  In agreement with
\cite{bow01}, we find that cooling seems to ``kick in" at a mass scale
slightly smaller than required by observations.  This discrepancy of mass
scales may indicate that some non-gravitational heating is still required.
Certainly high-redshift galaxies are observed to produce strong winds
\citep{pet01}, and AGN activity is sometimes seen in the centers of
groups, so potential sources of non-gravitational heating do exist.
Some recent models of cluster ``cooling flows'' incorporate an additional
heat source that serves to maintain the inner gas at an intermediate
temperature, suppressing the spectral signatures of $\sim 1$~keV gas that
would otherwise be expected \citep{david01,fab01}.  Hence, it is plausible
that baryons in groups have experienced some non-gravitational heating,
but the amount needed to reconcile models with observations may be much
less than previously proposed, thus alleviating the ``ICM energy crisis."

Fortunately, this field appears poised for breakthroughs on both
observational and theoretical fronts.  {\it XMM} and {\it Chandra} will
allow studies of groups with greater sensitivity, spatial resolution,
and spectral resolution than previously possible.  Constraints on optical
properties of groups continue to improve with deeper surveys.  On the
modeling side, much effort has gone into exploring solutions for the
``decoupling problem'' in two-phase regimes, with new algorithms being
proposed that may alleviate the problem in a self-consistent fashion
\citep[e.g.][]{tha00,spr01}.  Furthermore, Moore's Law improvement
in computing technology continues unabated, enabling ever larger and
higher resolution simulations.  Advances on both fronts will soon enable
reliable side-by-side comparisons of simulations and observations,
where the simulations are analyzed using exactly the same techniques as
the data.  This approach promises to yield an in-depth understanding of
the physical processes governing the formation of galaxies in their most
ubiquitous environment, groups.

\acknowledgments
We thank Arif Babul, Richard Bower, Mark Fardal, John Mulchaey,
David Spergel, Mark Voit, and Marc Zimer for useful discussions.
We thank Geraint Lewis and Chigurupati Murali for assembling the
Raymond-Smith code and TIPSY interface, and Jeff Gardner for 
use of his spherical overdensity group finder.
Support for this work was provided by NASA through grants
GO-08165.01-97A and NAG5-3525, by NSF Grant AST-9802568,
and by Hubble Fellowship Grant HST-HF-01128.01-A
from the Space Telescope Science Institute, which is operated by
AURA, Inc., under NASA contract NAS5-26555.

\end{document}